\documentclass[12pt,a4j]{article}
\usepackage[dvips]{graphicx}
\usepackage{amsmath}
\usepackage{amssymb}
\usepackage{enumerate}

\begin{document}
\baselineskip=7mm
\centerline{\bf  A direct method for solving the generalized sine-Gordon equation }\par
\bigskip
\centerline{Yoshimasa Matsuno}\par
\centerline{\it Division of Applied Mathematical Science, Graduate School of Science and Engineering} \par
\centerline{\it Yamaguchi University, Ube, Yamaguchi 755-8611, Japan} \par
\medskip
\centerline{{\it E-mail address}: matsuno@yamaguchi-u.ac.jp}\par
\bigskip
\bigskip
\noindent {\bf Abstract} \par
\noindent The generalized sine-Gordon (sG) equation was derived as an integrable
generalization of the sG equation. In this paper, we develop a direct method for solving the
generalized sG equation without recourse to the inverse scattering method.
 In particular, we construct multisoliton solutions in the form of parametric representation.
 We obtain a variety of solutions which include  kinks, loop solitons and breathers. 
 The properties of these solutions are investigated in detail.
  We find a novel type of solitons with a peculiar structure that  the smaller soliton
travels faster than the larger soliton. We also show that the short pulse equation describing
the propagation of ultra-short pulses is
reduced from the generalized sG equation in an appropriate scaling limit. 
Subsequently, the reduction to
the sG equation is briefly discussed.

\par

\bigskip
\bigskip
\leftline{{\bf MSC}: 35Q51, 37K10, 37K40}\par
\newpage
\leftline{\bf  1. Introduction} \par
\bigskip
\noindent We consider the following generalized sine-Gordon (sG) equation
$$u_{tx}=(1+\nu\partial_x^2)\sin\,u, \eqno(1.1)$$
where $u=u(x,t)$ is a scalar-valued function, $\nu$ is a real parameter, 
 $\partial_x^2=\partial^2/\partial x^2$ and the subscripts $t$ and $x$ appended to $u$ denote partial differentiation. 
 The generalized sG equation
has been derived for the first time in [1] using bi-Hamiltonian methods. Quite recently,
 equation (1.1) with $\nu<0$ was shown to be a completely integrable partial differential 
 equation (PDE) [2]. Indeed, constructing a Lax pair
associated with it, the initial value problem of the equation   was solved for
decaying initial data. In the process, the Riemann-Hilbert formalism was developed to obtain
eigenfunctions of  the Lax pair. 
Soliton solutions are obtainable in principle, but their derivation needs a very complicated
procedure.
Although some qualitative features of
traveling-wave solutions are discussed in a different context from the
Riemann-Hilbert formalism, explicit expressions of solutions are not
available as yet. \par
The purpose of this paper is to obtain exact solutions of equation (1.1) with $\nu<0$
and discuss their properties. We consider real and nonperiodic solutions.
 In our analysis, we take $\nu=-1$ without loss of generality. 
 The exact method of solution used here is the  bilinear transformation method 
 which is a very powerful tool in obtaining special solutions of soliton equations [3, 4].
 The method has wide applications ranging from continuous to discrete soliton equations.
 The central problem in the bilinear formalism is the construction of tau-functions
 which are introduced through dependent variable transformations.
 \par
 This paper is organized as follows. In section 2, we develop an exact method of solution.
 Specifically, we use a hodograph transformation to transform equation under consideration
 into a more tractable form.  The transformed equation is further put into a system of
 bilinear equations by introducing appropriate dependent variable transformations.
 We then construct explicit solutions of the bilinear equations by means of a standard
 procedure in the bilinear formalism. The  
 multisoliton  solutions are obtained in the form of parametric representation.
 In section 3, we describe properties of solutions.  First, we consider 1-soliton solutions
 which include kink and loop soliton solutions as well as a new type of multivalued functions.
 Throughout this paper we use the  term "soliton" as a generic name of elementary solutions such as
 kink, loop soliton and breather solutions. A novel feature of regular kink solutions is found which
 has never been seen in the sG kinks. Next, the 2-soliton solutions are discussed. We address the kink-kink and kink-loop 
 soliton solutions together with the 1-breather solution. Last, we explore the general multisoliton solutions.
 Our particular concern is the multikink solution for which the large-time asymptotic is derived and the associated formula
 for the phase shift is obtained. A recipe for constructing the multibreather solutions is briefly described.
 As  examples of the multisoliton solutions, we present a solution describing the interaction
 between a soliton and a breather as well as a 2-breather solution 
 which are reduced from the 3 and 4-soliton solutions, respectively.
 In section 4, we point out a close relationship between the generalized sG equation and the short pulse equation which
 models the propagation of ultra-short optical pulses.
  We show that the generalized sG equation is reduced to
 the short pulse equation by taking an appropriate scaling limit. 
 The parametric multiloop soliton solution of the short pulse equation presented in [5] is reproduced from the corresponding
 one for the generalized sG equation. The similar limiting procedure is also applied to the formula for the phase shift.
 Subsequently, the reduction of the generalized sG equation to the sG equation is discussed shortly.
 Section 5 is devoted to
 conclusion. In the appendix A, we show that the tau-functions for the multisoliton solutions obtained in section 3 satisfy
 a system of bilinear equations.  The proof is carried out  by means of
 an elementary method using various formulas for determinants. 
 In the appendix B, we derive the 1-soliton solutions by  an alternative method and demonstrate that they reproduce the corresponding
 1-soliton solutions obtained in section 3.
 \par
 \bigskip
 \leftline{\bf 2. Exact method of solution} \par
 \bigskip
 \leftline{\it 2.1. Hodograph transformation} \par
 \medskip
 \noindent We introduce  the new dependent variable $r$ in accordance with the relation
$$r^2=1+u_x^2,\qquad (r>0), \eqno(2.1)$$
to transform equation (1.1) with $\nu=-1$ into the form
$$ r_t+(r\,\cos\,u)_x=0. \eqno(2.2)$$
We then define the hodograph transformation $(x ,t) \rightarrow (y, \tau)$ by
$$dy=rdx-r\,\cos\,u\,dt, \qquad d\tau=dt. \eqno(2.3)$$ 
 The $x$ and $t$ derivatives are then rewritten  in terms of the $y$ and $\tau$ derivatives  as
 $${\partial\over\partial x}=r{\partial\over\partial y}, \qquad {\partial\over\partial t}
 ={\partial\over\partial \tau}-r\,\cos\,u\,{\partial\over\partial y}. \eqno(2.4)$$
 With the new variables $y$ and $\tau$, (2.1) and (2.2) are recast into the form
 $$r^2=1+r^2u_y^2, \eqno(2.5)$$
 $$\left({1\over r}\right)_\tau-(\cos\,u)_y=0, \eqno(2.6)$$
 respectively. Further reduction is possible if one defines the variable $\phi$ by
 $$u_y=\sin\,\phi, \qquad \phi=\phi(y,\tau),\quad \left(-{\pi\over 2}<\phi<{\pi\over 2},\ {\rm mod}\ 2\pi\right). \eqno(2.7)$$
  It follows from (2.5) and (2.7) that
  $$ {1\over r}=\cos\,\phi. \eqno(2.8)$$
 Substituting (2.7) and (2.8) into equation (2.6), we find
 $$\phi_\tau=\sin\,u. \eqno(2.9)$$ 
 If we eliminate the variable $\phi$ from (2.7) and (2.9), we obtain a single PDE for $u$
 $${u_{\tau y}\over \sqrt{1-u_y^2}}=\sin\,u. \eqno(2.10)$$
 Similarly, elimination of the variable $u$ gives a single PDE for $\phi$
 $${\phi_{\tau y}\over \sqrt{1-\phi_\tau^2}}=\sin\,\phi. \eqno(2.11)$$
 By inverting the relationship (2.4) and using  (2.8),  equation that determines the inverse mapping $(y,\tau) \rightarrow (x,t)$ is  found to be governed 
by the system of linear  PDEs for $x=x(y,\tau)$
$$x_y=\cos\,\phi, \eqno(2.12a)$$
$$x_\tau=\cos\,u. \eqno(2.12b)$$ 
 Note that the integrability of the system of equations (2.12) is assured by (2.7) and (2.9). \par
 A sequence of transformations described above are almost the same as those employed for solving the
 short pulse equation [5]. The underlying idea is to transform the orignal equation to the (possibly) integrable equation.
 In the case of the short pulse equation, the transformrd equation is the sG equation whereas in the present case,
 the corresponding equations are (2.10) and (2.11). The soliton solutions of the latter equations will be
 construcred here for the first time. Given $u$ and $\phi$, the most difficult problem is how to integrate the
 system of equations (2.12). This becomes the core part of the present paper  and will be resolved by Theorem 2.1. \par 
 \bigskip
 \leftline{\it 2.2. Bilinear formalism}\par
 \medskip
 \noindent Here, we develop a method for solving a system of PDEs (2.7) and (2.9). 
 We use the bilinear transformation method [3, 4].
 Let $\sigma$ and $\sigma^\prime$
 be solutions of the sG equation
 $$\sigma_{\tau y}=\sin\,\sigma, \qquad \sigma=\sigma(y,\tau),\eqno(2.13a)$$
 $$\sigma^\prime_{\tau y}=\sin\,\sigma^\prime, \qquad \sigma^\prime=\sigma^\prime(y,\tau). \eqno(2.13b)$$
 We then put
 $$u={1\over 2}(\sigma+\sigma^\prime), \eqno(2.14a)$$
 $$\phi={1\over 2}(\sigma-\sigma^\prime). \eqno(2.14b)$$
 In terms of $\sigma$ and $\sigma^\prime$, equations (2.7) and (2.9) can be written as
 $${1\over 2}(\sigma+\sigma^\prime)_y=\sin\,{1\over 2}(\sigma-\sigma^\prime), \eqno(2.15a)$$ 
  $${1\over 2}(\sigma-\sigma^\prime)_\tau=\sin\,{1\over 2}(\sigma+\sigma^\prime). \eqno(2.15b)$$
 It should be remarked that the system of PDEs (2.15) constitutes a B\"acklund transformation of the
 sG equation with a  B\"acklund parameter taken to be 1 [6]. 
 The real-valued solutions of the sG equations (2.13) can be put into the form [7-10]
 $$\sigma=2{\rm i}\,\ln{f^*\over f}, \eqno(2.16a)$$
 $$\sigma^{\prime}=2{\rm i}\, \ln{g^*\over g}, \eqno(2.16b)$$
 where $f^*$ and $g^*$ denote the complex conjugate of $f$ and $g$, respectively. The tau-functions $f$ and $g$
 play a central role in our analysis. 
 They are fundamental quantities in constructing solutions. For soliton solutions, they satisfy the following bilinear
 equations [7]:
 $$D_\tau D_yf\cdot f={1\over 2}(f^2-{f^*}^2), \eqno(2.17a)$$
 $$D_\tau D_yg\cdot g={1\over 2}(g^2-{g^*}^2), \eqno(2.17b)$$
 where the bilinear operators $D_\tau$ and $D_y$ are defined by
 $$D_\tau^mD_y^nf\cdot g=\left(\partial_\tau-\partial_{\tau^\prime} \right)^m
                 \left(\partial_y-\partial_{y^\prime} \right)^nf(\tau,y)g(\tau^\prime,y^\prime)|_{\tau^\prime=\tau,\,
                 y^\prime=y}, \qquad (m, n=0, 1, 2, ...).   \eqno(2.18)$$
 \par
 Now, we seek solutions of equations (2.7) and (2.9) of the form 
 $$u={\rm i}\,\ln{F^*\over F}, \eqno(2.19a)$$
 $$\phi={\rm i}\,\ln{G^*\over G}, \eqno(2.19b)$$
 where  $F$ and $G$ are new tau-functions.
 It turns out from (2.14), (2.16) and (2.19) that
 $$2u=\sigma+\sigma^\prime=2{\rm i}\, \ln{f^*g^*\over fg}=2{\rm i}\,\ln{F^*\over F}, \eqno(2.20a)$$
 $$2\phi=\sigma-\sigma^\prime=2{\rm i}\, \ln{f^*g\over fg^*}=2{\rm i}\,\ln{G^*\over G}. \eqno(2.20b)$$
 By taking into account  (2.20), we may assume the following relations among the tau-functions $f, g, F$ and $G$:
 $$F=fg, \eqno(2.21a)$$
  $$G=fg^*. \eqno(2.21b)$$
 The above expressions lead to an important relation
 $$F^*F=G^*G. \eqno(2.22)$$
 Substituting (2.19) into equations (2.7) and (2.9) and using (2.22), we see that
 $F$ and $G$ satisfy a system of bilinear equations
  $$D_yF^*\cdot F=-{1\over 2}(G^2-{G^*}^2), \eqno(2.23a)$$
   $$D_\tau G^*\cdot G=-{1\over 2}(F^2-{F^*}^2). \eqno(2.23b)$$
 Thus, the problem under consideration is reduced to obtain solutions of equations (2.23) subjected to the condition (2.22).
 After some trials, however, we  found that this procedure for constructing solutions is difficult to perform. 
  Hence, we employ an alternative approach.  To begin with, we impose the following bilinear 
  equations for $f$ and $g$ which turn out to be the starting point in our analysis
  $$D_yf\cdot g^*={1\over 2}(fg^*-f^*g), \eqno(2.24a)$$
  $$D_\tau f\cdot g={1\over 2}(fg-f^*g^*). \eqno(2.24b)$$
  With (2.24) at hand, the following proposition holds: \par
  \noindent {\bf Proposition 2.1.} {\it If $f$ and $g$ satisfy the bilinear equations (2.24), then the tau-functions $F$ and $G$
  defined by (2.21) satisfy the bilinear equations (2.23).} \par
  \noindent {\bf Proof.} First, we prove (2.23a). We substitute (2.21a) into the left-hand side of (2.23a) and rewrite it
  in terms of bilinear operator to obtain
  $$D_yF^*\cdot F=-(D_yf\cdot g^*)f^*g+(D_yf^*\cdot g)fg^*. \eqno(2.25)$$
  By virtue of (2.24a), the right-hand side of (2.25) becomes $-(1/2)\{(fg^*)^2-(f^*g)^2\}$ which is equal to
  the right-hand side of (2.23a) by (2.21b). The proof of (2.23b) can be done in the same way by using (2.24b).
  \hspace{\fill} $\square$ \par
  \bigskip
  \leftline{\it 2.3. Parametric representation}\par
  \medskip
  \noindent We demonstrate that the solution of  equation (1.1) with $\nu=-1$ admits a parametric representation.
  The following relation is crucial to integrate (2.12):\par
  \noindent{\bf Proposition 2.2.} {\it $\cos\,\phi$ is expressed in terms of $f$ and $g$ as}
  $$\cos\,\phi=1+\left(\ln{g^*g\over f^*f}\right)_y. \eqno(2.26)$$
  \noindent {\bf Proof.} Using (2.24a), one obtains
  \begin{alignat*}{1}
    \left(\ln{g^*g\over f^*f}\right)_y &=-{D_yf\cdot g^*\over fg^*}-{D_yf^*\cdot g\over f^*g} \\
                                      &={1\over 2}{(fg^*)^2+(f^*g)^2 \over f^*fg^*g}-1. \tag{2.27}
  \end{alignat*}
  On the other hand, it follows from (2.19b) and (2.21b) that
  \begin{alignat*}{1}
  \cos\,\phi &={1\over 2}\left({G\over G^*}+{G^*\over G}\right) \\
             &= {1\over 2}{(fg^*)^2+(f^*g)^2\over f^*fg^*g}. \tag{2.28}
  \end{alignat*}
  The relation (2.26)  follows immediately by comparing (2.27) and (2.28). \hspace{\fill}$\square$ \par
  Integrating  (2.12a) coupled with (2.26) by $y$, we obtain the expression of $x$
  $$x=y+ \ln{g^*g\over f^*f}+d(\tau), \eqno(2.29)$$
  where $d$ is an integration constant which depends generally on $\tau$. The expression (2.29) now
  leads to our main result: \par
  \noindent{\bf Theorem 2.1.} {\it The real-valued solution of equation (1.1) with $\nu=-1$ can be written by the
  parametric representation 
   $$u(y,\tau)={\rm i}\,\ln{f^*g^*\over fg}, \eqno(2.30a)$$
  $$x(y,\tau)=y+\tau + \ln{g^*g\over f^*f} + y_0, \eqno(2.30b)$$
  where the tau-functions $f$ and $g$  satisfy both (2.17) and (2.24) simultaneously 
  and $y_0$ is an arbitrary constant independent of $y$ and $\tau$}. \par
  \noindent{\bf Proof.} The expression (2.30a) for $u$ is a consequence of (2.19a) and (2.21a).
  To prove (2.30b), we substitute (2.29) into (2.12b) and obtain the relation
  $$\cos\,u=\left(\ln{g^*g\over f^*f}\right)_\tau+d^\prime(\tau). \eqno(2.31)$$
  The left-hand side of (2.31) can be expressed by $f$ and $g$ in view of (2.30a) whereas
  the right-hand side is modified by using (2.24b). After a few calculations, we find that
  most terms are cancelled, leaving  
  the equation $ d^\prime(\tau)=1$. Integrating this equation, one obtains   $d(\tau)=\tau+y_0$,
  which, substituted into (2.29),  gives the expression (2.30b) for $x$. \hspace{\fill}$\square$ \par 
  The parametric solution (2.30) would produce in general a multi-valued function as happned in the case of the short pulse
  equation [5]. To derive a criterion for single-valued functions, we calculate $u_x$ with use of  (2.7) and (2.8) and  obtain
  $$u_x=ru_y=\tan\,\phi. \eqno(2.32)$$
  Thus, if the inequality $-\pi/2<\phi<\pi/2\ ({\rm mod}\ \pi)$  holds for all $y$ and $\tau$, then $u$ becomes a regular function of $x$ and $t$.
  By virtue of the identity
  $${\rm i}\,\ln{G^*\over G}=2\,\tan^{-1}\left({{\rm Im}\, G\over {\rm Re}\, G}\right), \eqno(2.33)$$
  as well as the relation (2.21b), the above condition for regularity can be written as
  $$-1<{{\rm Im}(fg^*)\over {\rm Re}(fg^*)}<1. \eqno(2.34)$$
  In the case of 1-soliton solutions discussed in the next section, the above condition is
  found explicitly in terms of the parameters characterizing solutions. \par
  \bigskip
  \leftline{\it 2.4. Multisoliton solutions} \par
  \medskip
  \noindent The last step in constructing solutions is to find the tau-functions $f$ and $g$ for the
  sG equation which satisfy simultaneously the bilinear equations (2.24).  The following theorem
  establishes this purpose: \par
  \noindent{\bf Theorem 2.2.} {\it The tau-functions $f$ and $g$ given below satisfy both the bilinear forms (2.17) of the
  gG equation  and the
  bilinear equations (2.24)
  $$f=\sum_{\mu=0,1}{\rm exp}\left[\sum_{j=1}^N\mu_j\left(\xi_j+{\pi\over 2}\,{\rm i}\right)
+\sum_{1\le j<k\le N}\mu_j\mu_k\gamma_{jk}\right], \eqno(2.35a)$$
  $$g=\sum_{\mu=0,1}{\rm exp}\left[\sum_{j=1}^N\mu_j\left(\xi_j-2d_j+{\pi\over 2}\,{\rm i}\right)
+\sum_{1\le j<k\le N}\mu_j\mu_k\gamma_{jk}\right], \eqno(2.35b)$$
where
$$\xi_j=p_jy+{1\over p_j}\tau+\xi_{j0}, \qquad (j=1, 2, ..., N),\eqno(2.36a)$$
$${\rm e}^{\gamma_{jk}}=\left({p_j-p_k\over p_j+p_k}\right)^2, \qquad (j, k=1, 2, ..., N; j\not=k),
\eqno(2.36b)$$
$$e^{-2d_j}={1-p_j\over 1+p_j},\qquad (j=1, 2, ..., N).\eqno(2.36c)$$
Here, $p_j$ and $\xi_{j0}$ are arbitrary real parameters satisfying the conditions $p_j\not= p_k$
for $j\not= k$ and $N$ is an arbitrary positive integer. The notation $\sum_{\mu=0,1}$
implies the summation over all possible combination of $\mu_1=0, 1, \mu_2=0, 1, ..., 
\mu_N=0, 1$}.\par 
\noindent{\bf Proof.} It has been shown that $f$ and $g$ given by (2.35a) and (2.35b) satisfy the
bilinear equations (2.17a) and (2.17b), respectively [7]. Thus, it is sufficient to prove that they satisfy
the bilinear equations (2.24). The proof is carried out by a lengthy calculation using various formulas 
for determinants. It will be summarized in the appendix A. \hspace{\fill}$\square$ \par
\bigskip
\leftline{\it 2.5. Remark}\par
\medskip
\noindent The tau-functions $f$ and $g$ given by (2.35) yield real-valued solutions since all the
parameters $p_j$ and $\xi_{j0}\ (j=1, 2, ..., N)$ are chosen to be real numbers. If one looks for 
breather solutions, for example, one needs to introduce complex parameters (see sections 3.2.3, 3.3.2 and 3.3.3).
Even in this case, however, the analysis developed here can be applied as well without making essential changes.
Actually, we may use the tau-functions $f^\prime$ and $g^\prime$ instead of $f^*$ and $g^*$, respectively
where $f^\prime$ and $g^\prime$ are obtained simply from $f$ and $g$ by replacing ${\rm i}$ by ${\rm -i}$, but
all the parameters in the tau-functions are assumed to be complex numbers. 
The solutions of the sG  equations (2.13a) and (2.13b) can be written as $\sigma=2{\rm i}\,\ln(f^\prime/f)$ and
$\sigma^\prime=2{\rm i}\,\ln(g^\prime/g)$, respectively, where the tau-functions $f, f^\prime, g$ and $g^\prime$ satisfy
the following systems of bilinear equations
$$D_\tau D_yf\cdot f={1\over 2}(f^2-{f^\prime}^2), \qquad D_\tau D_yf^\prime\cdot f^\prime={1\over 2}({f^\prime}^2-f^2), \eqno(2.37a)$$
$$D_\tau D_yg\cdot g={1\over 2}(g^2-{g^\prime}^2), \qquad D_\tau D_yg^\prime\cdot g^\prime={1\over 2}({g^\prime}^2-g^2). \eqno(2.37b)$$
The bilinear equations corresponding to (2.24) are then given by
$$D_yf\cdot g^\prime={1\over 2}(fg^\prime-f^\prime g), \qquad
  D_\tau f\cdot g={1\over 2}(fg-f^\prime g^\prime), \eqno(2.38a))$$
$$D_yf^\prime\cdot g={1\over 2}(f^\prime g-fg^\prime), \qquad
  D_\tau f^\prime\cdot g^\prime={1\over 2}(f^\prime g^\prime-fg). \eqno(2.38b))$$
The expressions corresponding to (2.19)-(2.23) are obtained  if one  replaces the asterisk appended to the tau-functions by the prime.
Under these modifications, the real-valued solutions are
 produced if one imposes the conditions $f^\prime=f^*$ and $g^\prime=g^*$. 
 See also an analogous problem associated with breather solutions of the short pulse equation [5].
 \par
    \bigskip
      \leftline{\bf 3. Properties of solutions}\par
  \bigskip
  \noindent The parametric representation (2.30) of the solution with the tau functions given by (2.35) exhibit 
  a variety of soliton solutions of equation (1.1) with $\nu=-1$. As exemplified here,
 solutions include both single-valued and  multi-valued kinks, loop solitons and breathers. \par
\medskip
    \leftline{\it 3.1. 1-soliton solutions}\par
    \medskip
 \noindent  The tau-functions for the 1-soliton solutions are given by (2.35) with $N=1$:
  $$f=1+{\rm i}\,{\rm e}^{\xi_1}, \qquad \xi_1=p_1y+{\tau\over p_1}+\xi_{10}, \eqno(3.1a)$$
  $$g=1+{\rm i}s_1{\rm e}^{\xi_1},\qquad s_1={1-p_1\over 1+p_1}. \eqno(3.1b)$$
  The parameters $p_1$ and $\xi_{10}$ are related to the amplitude and phase of the soliton, respectively
  and $\xi_1$ is a phase variable characterizing the parametric representation of the solution.
  It follows from (2.21) and (3.1) that
  $$F=1-s_1{\rm e}^{2\xi_1}+{\rm i}\,(1+s_1){\rm e}^{\xi_1}, \eqno(3.2a)$$
  $$G=1+s_1{\rm e}^{2\xi_1}+{\rm i}\,(1-s_1){\rm e}^{\xi_1}. \eqno(3.2b)$$
  Using (2.30) and (3.1), the parametric representation of the solution is written in the form
  $$u=2\,\tan^{-1}(\sinh\,\xi_1-p_1\,\cosh\,\xi_1)+\pi, \eqno(3.3a)$$
  $$x=y+\tau+\ln\left({1+p_1^2\over (1+p_1)^2}-{2p_1\over (1+p_1)^2}\tanh\,\xi_1\right)+y_0, \eqno(3.3b)$$
  where we have imposed the boundary condition $u(-\infty,t)=0$.  
  Note that if $u$ solves equation (1.1), then so do the functions $\pm u+2\pi n\ (n : {\rm integer})$.
  To describe solutions of traveling-wave type like 1-soliton solutions, it is convenient to parameterize the solutions
   in terms of the single
  variable $\xi_1$.  For this purpose, we rewrite (3.3b)  as
  $$X\equiv x+c_1t+x_0={\xi_1\over p_1}+\ln\left({1+p_1^2\over (1+p_1)^2}-{2p_1\over (1+p_1)^2}\tanh\,\xi_1\right)+y_0, \eqno(3.4a)$$
   where $c_1$ is the velocity of the soliton given by
   $$c_1={1\over p_1^2}-1, \eqno(3.4b)$$
   and $x_0=\xi_{10}/p_1$. Observing the motion of the soliton in the original $(x,t)$ coordinate system,  
   it travels to the left at the constant velocity $c_1$ 
   for $p_1<1$ and to the right for $p_1>1$. In the critical case $p_1=1$ for which $c_1=0$, the soliton remains stationary.
  The profile of the soliton changes drastically depending on values of the parameter $p_1$. The singular nature of the solution can be extracted 
  conveniently from the information on   the gradient of $u$ with respect to $X$.  A calculation using (3.3a) and (3.4) gives
  $$u_X={1\over p_1\,\cosh\,\xi_1}{1-p_1\,\tanh\,\xi_1\over \tanh^2\xi_1-{1\over p_1}\tanh\,\xi_1+{1-p_1^2\over 2p_1^2}}. \eqno(3.5)$$
  \par
  Let us now analyze the solution (3.3).   Using (3.1), the condition (2.34) for single-valued function becomes
  $$-1<{p_1\over \cosh\,\xi_1-p_1\,\sinh\,\xi_1}<1. \eqno(3.6)$$ 
  In the following analysis, we assume $p_1>0$ without loss of generality.  
  If $0<p_1<1$, then (3.6) leads to the inequality
  $$\tanh^2\xi_1-{1\over p_1}\tanh\,\xi_1+{1-p_1^2\over 2p_1^2}>0. \eqno(3.7)$$
  One can see that the inequality (3.7) always holds for $0<p_1<{1\over \sqrt{2}}$. In this case, $u_X$ from (3.5) becomes
  finite for arbitrary values of $\xi_1$. On the other hand, if ${1\over \sqrt{2}}<p_1<1$, then the solution
  exhibits singularities at two points $X= X_\pm \equiv X(\xi_\pm)$ where
  $$\xi_{\pm}=\tanh^{-1}\left[{1\over 2p_1}\left(1\pm \sqrt{2p_1^2-1}\right)\right]. \eqno(3.8)$$
  If $p_1>1$, the inequality (3.6) breaks down, giving rise to two singular points whose positions are
  the same as $X_\pm$.
  Unlike the second case, however, $u_X$ becomes zero at $X=X_0\equiv X(\xi_0)$ with $\xi_0=\tanh^{-1}(1/p_1)$, as readily
  noticed from (3.5). 
  The solution for the case $p_1=1$ exhibits a peculiar behavior, which deserves a separate study.
  In view of
  these observations, the solutions can be classified to
  four types according to values of $p_1$, or equivalently $c_1$ by (3.4b), which we shall now investigate in detail.\par
  \medskip
  \noindent {\it Type 1. Regular kink: \ $c_1>1\ (0<p_1<{1\over \sqrt{2}}) $} \par
  \noindent  Figure 1 shows a typical profile of $u$ as a function of $X$. 
       It exhibits a profile of a $2\pi$-kink similar
   to the kink solution of the sG equation [6]. The propagation characteristic is, however, different from that of the sG kink. 
   To clarify this point, we rewrite $u_X$ from (3.5) as
   $$u_X={2p_1\over \sqrt{1-p_1^2}}{\cosh(\xi_1-d_1)\over \cosh^2(\xi_1-d_1)-{p_1^2\over 1-p_1^2}}, \eqno(3.9)$$
   where $d_1=(1/2)\ln[(1+p_1)/(1-p_1)]$. It turns out from (3.9) that the amplitude $A$ of $u_X$ is given by
   $$A={2p_1 \sqrt{1-p_1^2}\over 1-2p_1^2}. \eqno(3.10)$$
   Eliminating $p_1$ from (3.4b) and (3.10), we find the dependence of the amplitude
   on the velocity
   $$c_1={1\over A^2}\left(A^2+2+2\sqrt{A^2+1}\right). \eqno(3.11)$$
   The relation (3.11) indicates that $c_1$ is a monotonically decreasing function of $A$. In other words, the smaller soliton
   travels faster than the larger soliton. This peculiar feature of the solution  has never been observed in the
   behavior of the sG kink solutions.  
    The broken line in figure 1 plots the profile of $v\equiv u_X$ obtained from the kink solution depicted in the same
    figure.  It represents a soliton.
    \par
    \medskip
   \centerline{\bf Figure 1} \par
    \medskip
    \leftline{\it Type 2. Singular kink: \ $0<c_1<1\ ({1\over \sqrt{2}}<p_1<1)$}\par
    \noindent The profile of the solution of type 2 is a $2\pi$ kink, but in the interval $X_-<X<X_+$, it becomes a three-valued function.
    Figure 2 shows a typical profile of $u$ as a function of $X$. \par
   \medskip
   \centerline{\bf Figure 2} \par
   \medskip
   \leftline{\it Type 3. Loop soliton: \ $-1<c_1<0\ (p_1>1)$} \par
  \noindent When compared with solutions of types 1 and 2, the solution for $p_1>1$ exhibits a different behavior. Indeed, we see from
   (3.3), and (3.5)  that $u(\pm\infty)=0, u_X(X_0)=0$, respectively.  In addition, the solution has two
   singular points at $X=X_\pm$ such that that $X_0=(X_++X_-)/2$. Figure 3 shows a typical profile of $u$ as a function of $X$.
   It represents a loop soliton and has a symmetrical profile with respect to a straight line $X=X_0$. \par
     \medskip
   \centerline{\bf Figure 3} \par
   \medskip
    \leftline{\it Type 4. Stationary solution: \ $c_1=1\ (p_1=0)$} \par
   \noindent For this special value of the parameter $p_1$, the parametric 
   solution (3.3a) and (3.4a) takes the form
   $$u=-2\,\tan^{-1}{\rm e}^{-\xi_1}+\pi,\eqno(3.12a)$$
   $$X=-\ln\,({\rm e}^{\xi_1}+{\rm e}^{-\xi_1}). \eqno(3.12b)$$
   Remarkably, one can eliminate the variable $\xi_1$ from this expression to give an explicit solution
   $$\sin\,u=2\,{\rm e}^X, \qquad (-\infty<X<-\ln\,2). \eqno(3.13)$$
    The profile of $u$ is illustrated in figure 4 as a function of $X$.    
    Since $c_1=0$, one has $X=x+x_0$ by (3.4a) and the solution becomes time-independent. Keeping this fact in mind, we can 
    derive the solution (3.13)
   directly from the stationary version of equation (1.1) with $\nu=-1$
   $$\left(1-{d^2\over dx^2}\right)\sin\,u=0. \eqno(3.14)$$
   Integrating equation (3.14) under the boundary condition $u(-\infty)=0\ ($\rm mod$\, \pi)$,  we recover (3.13). \par
    \medskip
   \centerline{\bf Figure 4} \par
      \bigskip
      The 1-soliton solutions presented above are of fundamental importance in constructing general
      multisoliton solutions. In fact, the latter solutions will be shown to consist of any combination of
      the former solutions. In appendix B, we shall derive the 1-soliton solutions by means of an elementary
      method. \par
      \medskip
   \leftline{\it 3.2. 2-soliton solutions}\par
   \medskip
   \noindent The tau-functions  for the 2-soliton solutions read from (2.35) with $N=2$ in the form
   $$ f=1+{\rm i}({\rm e}^{\xi_1}+{\rm e}^{\xi_2})-\delta {\rm e}^{\xi_1+\xi_2}, \eqno(3.15a)$$
   $$ g=1+{\rm i}(s_1{\rm e}^{\xi_1}+s_2{\rm e}^{\xi_2})-\delta s_1s_2{\rm e}^{\xi_1+\xi_2}, \eqno(3.15b)$$
   where $\delta=(p_1-p_2)^2/(p_1+p_2)^2$ and $s_i=(1-p_i)/(1+p_i)\ (i=1, 2).$
   \par
   The parametric solution (2.30) represents various types of solutions describing the interaction of two solitons. 
   Here, soliton means one of four types of solutions given in section 3.1.
   In addition, the solution exhibits a breather solution when the two parameters $p_1$ and $p_2$ appear as a complex conjugate pair.
   Although a number of solutions yield according to the combination of elementary solutions, we present three types 
   of solutions, i.e., kink-kink (or two-kink) solution and
   kink-loop soliton solution and a breather solution. \par
   \bigskip
   \leftline{\it 3.2.1. Kink-kink solution}\par
   \medskip
   \noindent The solid line in figure 5 a-c exhibits the profile of solution $u$  for three different times. 
   The solution represents the $4\pi$-kink.
      In the same figure, we depict $v\equiv u_x$ by the broken line, showing that it represents the interaction of two solitons. The characteristic of
    the interaction process of two solitons is seen to be quite different from that of the sG two solitons. 
    Indeed, as evidenced from figure 5, a smaller soliton overtakes,
    interacts and emerges ahead of a larger soliton. 
    This reflects the fact that the velocity of each soliton is a monotonically
    decreasing function of its amplitude (see (3.11)). 
        After the interaction, both solitons suffer phase shifts.
    The general formula for the phase shift arising from the interaction of $N$ solitons will be given by (3.26) below.
    In particular, for $N=2$, it reads
    $$\Delta_1=-{1\over p_1}\ln\left({p_1-p_2\over p_1+p_2}\right)^2-\ln\left({1+p_2\over 1-p_2}\right)^2, \eqno(3.16a)$$
     $$\Delta_2={1\over p_2}\ln\left({p_1-p_2\over p_1+p_2}\right)^2+\ln\left({1+p_1\over 1-p_1}\right)^2. \eqno(3.16b)$$
     In the present example, formula (3.16) yields $\Delta_1=4.55$ and $\Delta_2=-2.42.$ 
     A careful inspection of (3.16) reveals that $\Delta_1>0$ and$\Delta_2<0$ for arbitrary values of $p_1$ and $p_2$
      satisfying the inequality $0<p_1<p_2<1/\sqrt{2}$.
      This implies that the small soliton has moved forward and the large soliton backward 
     relative to the positions they would have reached if both solitons had moved at constant velocities throughout the interaction process.
       The novel feature of the 2-soliton solution described above will appear here for the
    first time.    
    \par
    \medskip
   \centerline{\bf Figure 5 a-c} \par
   \bigskip
   \leftline{\it 3.2.2. Kink-loop soliton solution}\par
   \medskip 
   \noindent The next example is a solution $u$ representing the interaction of a $2\pi$-kink and a loop soliton. See figure 6 a-c.
   Since the kink propagates to the left and the loop soliton to the right, the solution describes the head-on collision unlike 
   the first  example which exhibits an overtaking collision. \par
   \medskip
   \centerline{\bf Figure 6 a-c} \par
      \bigskip
   \leftline{\it 3.2.3. Breather solution}\par
   \medskip
   \noindent The breather solution can be interpreted as a bound state composed of a kink and antikink pair  in the
   sG model [6]. 
   It has a localized structure which oscillates with time and decays exponentially in space.
   In the generalized sG equation, the similar breather solutions to the sG breathers will be  shown to exist.
   The procedure for constructing breather solutions is the same as that has been used for the short pulse equation [5].
   To be more specific, let
   $$p_1=a+{\rm i}b, \qquad p_2=a-{\rm i}b=p_1^*,\qquad (a>0,\ b>0), \eqno(3.17a)$$
   $$\xi_{10}=\lambda+{\rm i}\mu, \qquad \xi_{20}=\lambda-{\rm i}\mu=\xi_1^*, \qquad (\lambda,\ \mu:\ {\rm real}). \eqno(3.17b)$$
   Then, $f$ and $g$ from (3.5) become
   $$ f=1+{\rm i}({\rm e}^{\xi_1}+{\rm e}^{\xi_1^*})+\left({b\over a}\right)^2{\rm e}^{\xi_1+\xi_1^*}, \eqno(3.18a)$$
   $$ g=1+{\rm i}(s_1{\rm e}^{\xi_1}+s_1^*{\rm e}^{\xi_1^*})+s_1s_1^*\left({b\over a}\right)^2{\rm e}^{\xi_1+\xi_1^*}, \eqno(3.18b)$$
      where 
      $$\xi_1=\theta+{\rm i}\chi, \eqno(3.18c)$$
$$\theta=a\left(y+{1\over a^2+b^2}\tau\right)+\lambda, \eqno(3.18d)$$
$$\chi=b\left(y-{1\over a^2+b^2}\tau\right)+\mu, \eqno(3.18e)$$
$$s_1={1-a^2-b^2-2{\rm i}b\over (1+a)^2+b^2}\equiv \alpha{\rm e}^{-{\rm i}\beta}. \eqno(3.18f)$$ 
One can rewrite $f$ and $g$ in terms of the  new variables defined by (3.18) as
   $$f=1 +\left({b\over a}\right)^2{\rm e}^{2\theta}+2{\rm i}\,{\rm e}^{\theta}\cos\,\chi, \eqno(3.19a)$$
   $$g=1 +\alpha^2\left({b\over a}\right)^2{\rm e}^{2\theta}+2{\rm i}\alpha {\rm e}^{\theta}\cos\,(\chi-\beta). \eqno(3.19b)$$
         The regular solution is obtainable if the inequality (2.34) holds with $f$ and $g$ being given by (3.19).
         An inspection reveals that if $a/b$ is sufficiently small compared to 1, then the solution would
         exhibit no singularities. However, it is not easy to extract the condition for the regularity from (2.34)
         when compared with the corresponding problem for the breather solution of the short pulse equation [5].
         We leave it to a future work. Instead, we present a regular solution by a numerical example.
               In figure 7a-c, the profile of $v\equiv u_x$ is depicted for three different times. We can observe that the breather propagates to
   the left while changing its profile. The propagation characteristic of the breather is similar to that of the short pulse equation [5]. \par
      \medskip
   \centerline{\bf Figure 7a-c} \par
      \bigskip
        \leftline{\it 3.3. $N$-soliton solutions} \par
        \medskip
   \noindent The solutions including an arbitrary number of solitons can be constracted from (2.30) with the tau-functions (2.35). There exist 
   a variety of solutions which are composed of any combination of 1-soliton solutions presented in section 3.1. 
  Here, we address the $N$-kink solutions and $M(=2N)$ breather solutions. 
    For the former solutions, we investigate the asymptotic behavior of solutions for
   large time and derive the formulas for the phase shift while for the latter ones, we provide a recipe for
   constructing $M$ breather solution from the $N$-soliton solution. As  examples, we present a 
   solution describing the interaction between a soliton and a breather as well as a two-breather solution. \par 
      \par
   \bigskip
   \leftline{\it 3.3.1.  N-kink solution} \par
   \medskip
   \noindent Let the velocity of the $j$th kink be $c_j=(1/p_j^2)-1, (0<p_j<1/\sqrt{2})$ 
   and order the magnitude of the velocity of each kink
as $c_1>c_2> ...>c_N$.  We observe the interaction of $N$ kinks in a moving frame
with a constant velocity $c_n$. We  take the limit $t \rightarrow -\infty$
with the phase variable $\xi_n$ being fixed.  We then find that $f$ and $g$
have the following leading-order asymptotics
$$ f \sim \delta_{n}\,\exp\left[\sum_{j=n+1}^N\left(\xi_j+{\pi\over 2}{\rm i}\right)\right]
\left(1+{\rm i}{\rm e}^{\xi_n+\delta_n^{(-)}}\right), \eqno(3.20a)$$
$$ g \sim \delta_{n}\,\exp\left[\sum_{j=n+1}^N\left(\xi_j-2d_j+{\pi\over 2}{\rm i}\right)\right]
\left(1-{\rm i}{\rm e}^{\xi_n-2d_n+\delta_n^{(-)}}\right), \eqno(3.20b)$$
where
$$\delta_n^{(-)}=\sum_{j=n+1}^N\ln\left({p_n-p_j\over p_n+p_j}\right)^2, \eqno(3.20c)$$
$$\delta_{n}=\prod_{n+1\leq j<k\leq N}\left({p_j-p_k\over p_j+p_k}\right)^2. \eqno(3.20d)$$
If we substitute (3.20) into (2.30), we obtain the asymptotic form of $u$ and $x$:

$$u \sim 2\,\tan^{-1}\left[{\sinh(\xi_n-d_n+\delta_n^{(-)})\over \cosh\,d_n}\right]+\pi, \eqno(3.21a)$$
$$x \sim y+\tau-\ln{1+p_n\tanh(\xi_n-d_n+\delta_n^{(-)})\over 1-p_n\tanh(\xi_n-d_n+\delta_n^{(-)})}
-4\sum_{j=n+1}^Nd_j-2d_n+y_0. \eqno(3.21b)$$
Note that we have used equivalent but different expressions for $u$ and $x$ from those given by (3.3).
\par
As $t \rightarrow +\infty$, the expressions corresponding to (3.21) are given by
$$u \sim 2\,\tan^{-1}\left[{\sinh(\xi_n-d_n+\delta_n^{(+)})\over \cosh\,d_n}\right]+\pi, \eqno(3.22a)$$
$$x \sim y+\tau-\ln{1+p_n\tanh(\xi_n-d_n+\delta_n^{(+)})\over 1-p_n\tanh(\xi_n-d_n+\delta_n^{(+)})}
-4\sum_{j=1}^{n-1}d_j-2d_n+y_0. \eqno(3.22b)$$
with
$$\delta_n^{(+)}=\sum_{j=1}^{n-1}\ln\left({p_n-p_j\over p_n+p_j}\right)^2. \eqno(3.22c)$$
\par
Let $x_c$ be the center position of the $n$th kink in the $(x,t)$ coordinate
system. It simply stems from the relations $\xi_n-d_n+\delta_n^{(\pm)}=0$ by invoking (3.21a)
and (3.22a). Thus, as  $t \rightarrow -\infty$
$$x_c+c_nt+x_{n0} \sim {1\over p_n}(d_n-\delta_n^{(-)})  -4\sum_{j=n+1}^Nd_j-2d_n+y_0, \eqno(3.23)$$
where $x_{n0}=\xi_{n0}/p_n$.
 As $t \rightarrow +\infty$, on the other hand,
the corresponding expression turns out to be 
$$x_c+c_nt+x_{n0} \sim {1\over p_n}(d_n-\delta_n^{(+)})  -4\sum_{j=1}^{n-1}d_j-2d_n+y_0, \eqno(3.24)$$
If we take into account the fact that all kinks propagate to the left, we can
define the phase shift of the $n$th kink as
$$\Delta_n=x_c(t\rightarrow -\infty)-x_c(t\rightarrow +\infty). \eqno(3.25)$$
Using (2.36c), (3.20c), (3.22c), (3.23) and (3.24), we find that
$$\Delta_n={1\over p_n}\left\{\sum_{j=1}^{n-1}\ln\left({p_n-p_j\over p_n+p_j}\right)^2
-\sum_{j=n+1}^N\ln\left({p_n-p_j\over p_n+p_j}\right)^2\right\}$$
$$+\sum_{j=1}^{n-1}\ln\left({1+p_j\over 1-p_j}\right)^2-\sum_{j=n+1}^{N}\ln\left({1+p_j\over 1-p_j}\right)^2,
\quad (n = 1, 2, ..., N). \eqno(3.26)$$
The first term on the right-hand side of (3.26)  coincides with the formula
for the phase shift arising from the interaction of $N$ kinks of the
sG equation [7, 8, 10] whereas the second and third terms appear as a consequence of the coordinate
transformation (2.3). \par
    \bigskip
   \leftline{\it 3.3.2.  M-breather solution} \par
   \medskip
   \noindent  The construction of the $M$-breather solution can be done following the similar procedure  to that for the
   1-breather solution developed in section 3.2.3. Here, $M(=2N)$ is a positive even number.
    To proceed, we specify the parameters in (2.35) and (2.36) for the tau-functions $f$ and $g$ as
$$p_{2j-1}=p_{2j}^*\equiv a_j+{\rm i}b_j,\quad a_j>0,\quad b_j>0,
\quad (j=1, 2, ..., M),\eqno(3.27a)$$
$$\xi_{2j-1,0}=\xi_{2j,0}^*\equiv \lambda_j+{\rm i}\mu_j,\quad (j=1, 2, ..., M),\eqno(3.27b)$$
where $a_j$ and $b_j$ are positive parameters and $\lambda_j$ and $\mu_j$ are real parameters, respectively.
Then, the phase variables $\xi_{2j-1}$ and $\xi_{2j}$ are written as
$$\xi_{2j-1}=\theta_j+{\rm i}\chi_j, \quad (j=1, 2, ..., M),\eqno(3.28a)$$
$$\xi_{2j}=\theta_j-{\rm i}\chi_j, \quad (j=1, 2, ..., M),\eqno(3.28b)$$
with
$$\theta_j=a_j(y+c_j\tau)+\lambda_j, \quad (j=1, 2, ..., M), \eqno(3.28c)$$
$$\chi_j=b_j(y-c_j\tau)+\mu_j, \quad (j=1, 2, ..., M), \eqno(3.28d)$$
$$c_j={1\over a_j^2+b_j^2}, \quad (j=1, 2, ..., M). \eqno(3.28e)$$
The parametric solution (2.30) with (3.27) and (3.28) describes
multiple collisions of $M$  regular breathers provided that certain condition is
imposed on the parameters $a_j$ and $b_j (j=1, 2, ..., M)$. 
Although  it will be a  difficult task to derive the condition for the regularity through the  inequality  (2.34),
 numerical examples confirm the existence of regular multibreather solutions.
 See also an example of the 2-breather solution of the short pulse equation [5].
The asymptotic analysis for the $M$-breather solution can be performed following the procedure for 
   the $N$-kink solution, showing that the $M$-breather solution splits into $M$ single
   breathers as $t\rightarrow \pm\infty$.
     The resulting asymptotic form is, however,  too complicated to write down and hence we omit the detail.
   One can refer to the similar analysis to that for the $M$-breather solution of the short pulse equation [5].\par
   \bigskip
   \leftline{\it 3.3.3 Soliton-breather solution}\par
   \medskip
   \noindent  We take a 3-soliton solution with parameters  $p_j$ and $\xi_{0j}$ $(j=1, 2, 3)$.
   If one impose the conditions $p_2=p_1^*, \xi_{02}=\xi_{01}^*$ as  already specified for the breather solution (see section 3.2.3) 
   and $0<p_3<1/\sqrt{2},\xi_{03}$(:real) for the regular kink solution, then
   the expression of $v\equiv u_x$ would represent a solution describing the interaction between a soliton and a breather.
   We choose $p_1, p_2, \xi_{10}$ and $\xi_{20}$ as those given by (3.17). Then, the tau-functions $f$ and $g$ from (2.35)
   become
   $$f=1+{\rm i}({\rm e}^{\xi_1}+{\rm e}^{\xi_1^*}+{\rm e}^{\xi_3})+\left({b\over a}\right)^2{\rm e}^{\xi_1+\xi_1^*}
   -\delta_{13}{\rm e}^{\xi_1+\xi_3}-\delta_{13}^*{\rm e}^{\xi_1^*+\xi_3}
   +{\rm i}\left({b\over a}\right)^2\delta_{13}\delta_{13}^*{\rm e}^{\xi_1+\xi_1^*+\xi_3}, \eqno(3.29a)$$
   $$g=1+{\rm i}(s_1{\rm e}^{\xi_1}+s_1^*{\rm e}^{\xi_1^*}+s_3{\rm e}^{\xi_3})+\left({b\over a}\right)^2s_1s_1^*{\rm e}^{\xi_1+\xi_1^*}
   -\delta_{13}s_1s_3{\rm e}^{\xi_1+\xi_3}-\delta_{13}^*s_1^*s_3{\rm e}^{\xi_1^*+\xi_3}$$
   $$+{\rm i}\left({b\over a}\right)^2\delta_{13}\delta_{13}^*s_1s_1^*s_3{\rm e}^{\xi_1+\xi_1^*+\xi_3}, \eqno(3.29b)$$
   where
   $$s_1={1-a-{\rm i}b\over 1+a+{\rm i}b}=s_2^*,\qquad s_3={1-p_3\over 1+p_3},
   \qquad \delta_{13}=\left({a-p_3+{\rm i}b\over a+p_3+{\rm i}b}\right)^2=\delta_{23}^*. \eqno(3.29c)$$
   \par
   Figure 8a-c  shows a profile of $v\equiv u_x$ for three different times. 
  We see that as time evolves the soliton overtakes the breather whereby it suffers a phase shift.
  An asymptotic analysis using the tau-functions (3.29) leads to the formula for the phase
  shift of the soliton, which we denote $\Delta$. Actually,  one has for $p_3^2<a^2+b^2$
  $$\Delta={2\over p_3}\, \ln{(p_3+a)^2+b^2\over (p_3-a)^2+b^2} -2\,\ln{(1+a)^2+b^2\over (1-a)^2+b^2}. \eqno(3.30a)$$
  and for $a^2+b^2<p_3^2$
  $$\Delta=-{2\over p_3}\, \ln{(p_3+a)^2+b^2\over (p_3-a)^2+b^2} +2\,\ln{(1+a)^2+b^2\over (1-a)^2+b^2}. \eqno(3.30b)$$
        In the present example, formula (3.30a) gives $\Delta=3.07$. \par
      \medskip
   \centerline{\bf Figure 8a-c}\par
   \bigskip
   \leftline{\it 3.3.4 Breather-breather solution}\par
   \noindent The breather-breather (or 2-breather) solution is reduced from a 4-soliton solution following the procedure
   described in section 3.3.2. Figure 9a-c shows  a profile of $v\equiv u_x$ for three different times. 
   It represents a typical feature common to the interaction of solitons, i.e., each breather recovers its profile
   after collision.  
   \par
    \medskip
   \centerline{\bf Figure 9a-c}\par
   \bigskip

    \bigskip
   \leftline{\bf 4. Reduction to the short pulse  and sG equations} \par
   \bigskip
   \noindent The short pulse  equation was proposed as a model nonlinear equation 
describing the propagation of ultra-short optical pulses in nonlinear media [11]. It
 may be written in an appropriate dimensionless form as
$$u_{tx}=u+{1\over 6 }(u^3)_{xx}, \eqno(4.1)$$
where $u=u(x,t)$ represents the magnitude of the electric field. Here, we demonstrate that the generalized
sG equation is reduced to the short pulse equation by taking an appropriate scaling limit combined with a
coordinate transformation. The $N$-soliton solution of the short pulse equation as well as the formula of
the phase shift can be derived from those of the generalized sG equation. 
The reduction to the SG equation is shown to be established as well by  another scaling limit.
\par
\bigskip
\leftline{\it 4.1. Reduction to the short pulse equation}\par
\medskip
 \noindent Let us first introduce new variable $\bar u, \bar x$ and $\bar t$ according to
the relations
$$\bar u={u\over \epsilon},\qquad \bar x={1\over\epsilon}(x-t),\qquad \bar t=\epsilon t, \eqno(4.2)$$ 
   where $\epsilon$ is a small parameter and the quantities with bar are assumed to be order 1. Rewriting the
   derivatives in terms of the new variables $\bar t$ and $\bar x$ and expanding $\sin\,\epsilon \bar u$ in an
   infinite series with respect to $\epsilon$, we can develop equation (1.1) with $\nu=-1$ to 
      $$\epsilon\left(\bar u_{\bar t\bar x}-{1\over \epsilon^2}\bar u_{\bar x\bar x}\right)
   =\epsilon \bar u-{ \epsilon^3\over 6}\bar u^3+ ...
   -{1\over \epsilon^2}\partial^2_{\bar x}\left(\epsilon \bar u 
   -{ \epsilon^3\over 6} \bar u^3+{\epsilon^5\over 120}\bar u^5 + ...\right). \eqno(4.3a)$$
   Note that the terms of order $\epsilon^{-1}$ are canceled. Thus, we have
   $$\epsilon\bar u_{\bar t\bar x}=\epsilon\left(\bar u+{1\over 6}(\bar u^3)_{\bar x\bar x})\right)+O(\epsilon^3). \eqno(4.3b)$$
   If we divide both sides of (4.3b) by $\epsilon$ and then take the limit $\epsilon\rightarrow 0$, 
   we arrive at the short pulse equation (4.1) written by the new variables.
   It is noteworthy that the key relation (2.1) has been used to transform the short pulse equation into
   the sG equation and it is invariant under the scaling (4.2). The similar scaling variables to (4.2)
   have been used to derive the short-wave models of the Camassa-Holm and Degasperis-Procesi equations [12].
   \par
   \bigskip
   \leftline{\it 4.1.1. Scaling limit of the N-soliton solution}\par
   \medskip
   \noindent In order to perform the scaling limit of the $N$-soliton solution, we find it appropriate to
   employ the following new variables in addition to the
   variables defined by (4.2)
   $$\bar y={y\over \epsilon} \qquad  \bar y_0={y_0\over\epsilon}, \qquad \bar\tau=\epsilon \tau,
   \qquad \bar p_j=\epsilon p_j, \qquad \bar\xi_{j0}= \xi_{j0},\quad (j=1, 2, ..., N). \eqno(4.4)$$
   If one rewrites the tau-function $f$ from (2.35a) in terms of these variables, one simply has
      $$f=\bar f\equiv \sum_{\mu=0,1}{\rm exp}\left[\sum_{j=1}^N\mu_j\left(\bar\xi_j+{\pi\over 2}\,{\rm i}\right)
+\sum_{1\le j<k\le N}\mu_j\mu_k\bar\gamma_{jk}\right], \eqno(4.5a)$$
   with
   $$\bar\xi_j=\bar p_j\bar y+{\bar\tau\over\bar p_j}+\bar\xi_{j0}, \qquad (j=1, 2, ..., N),\eqno(4.5b)$$
   $${\rm e}^{\bar\gamma_{jk}}=\left({\bar p_j-\bar p_k\over \bar p_j+\bar p_k}\right)^2, \qquad (j, k=1, 2, ..., N; j\not=k).
\eqno(4.5c)$$
   In proceeding to the limiting procedure for the tau-function $g$, we need to retain terms up to order $\epsilon$.
   Thus, the expansion
   $${\rm exp}\left(-2\sum_{j=1}^N\mu_jd_j\right)=\prod_{j=1}^N\left({1-p_j\over 1+p_j}\right)^{\mu_j}
   \sim {\rm exp}\left(-\pi {\rm i}\sum_{j=1}^N\mu_j\right)\left(1-2\epsilon \sum_{j=1}^N{\mu_j\over \bar p_j}\right)+O(\epsilon^2), \eqno(4.6)$$
   as well as the scaled variables (4.2) and (4.4) are substituted into (2.35b) to derive the expansion of $g$
      $$g \sim \sum_{\mu=0, 1}\left(1-2\epsilon \sum_{j=1}^N{\mu_j\over \bar p_j}\right)
   {\rm exp}\left[\sum_{j=1}^N\mu_j\left(\bar\xi_j-{\pi\over 2}{\rm i}\right)
+\sum_{1\le j<k\le N}\mu_j\mu_k\bar\gamma_{jk}\right]+O(\epsilon^2)$$
$$=\bar f^*-2\epsilon\bar f_{\bar\tau}^*+O(\epsilon^2). \eqno(4.7)$$
Insertion of (4.2), (4.5) and (4.7) into (2.30a) gives
$$\epsilon\bar u={\rm i}\,\ln{\bar f^*(\bar f-2\epsilon \bar f_{\tau})\over \bar f(\bar f^*-2\epsilon \bar f_{\tau}^*)}+O(\epsilon^2). \eqno(4.8)$$
In the limit of $\epsilon\rightarrow 0$, (4.8) leads to the scaling limit of $u$
$$\bar u=2{\rm i}\left(\ln\,{\bar f^*\over \bar f}\right)_{\bar \tau}. \eqno(4.9)$$
Applying the similar procedure to (2.30b), we find
$$\epsilon\bar x=\epsilon\bar y-2\epsilon(\ln\,\bar f^* \bar f)_{\bar\tau}+\epsilon \bar y_0+O(\epsilon^2). \eqno(4.10)$$
It follows from (4.10) that 
$$\bar x=\bar y-2(\ln\,\bar f^* \bar f)_{\bar\tau}+\bar y_0. \eqno(4.11)$$
The expressions (4.9) and (4.11) coincide with the parametric representation of the $N$-soliton solution of the
short pulse equation [5]. \par
\bigskip
\leftline{\it 4.1.2. Scaling limit of the phase shift}\par
\medskip
\noindent The scaling limit of the formula (3.26) for the phase shift can be derived easily. Indeed, if we define the new variable
for the phase shift by $\bar \Delta_n=\Delta_n/\epsilon$  and substitute this expression and the scaled 
variable $\bar p_j=\epsilon p_j$ from (4.4) into (3.26), we find,
after taking the limit $\epsilon\rightarrow 0$, that
$$\bar\Delta_n={1\over \bar p_n}\left\{\sum_{j=1}^{n-1}\ln\left({\bar p_n-\bar p_j\over \bar p_n+\bar p_j}\right)^2
-\sum_{j=n+1}^N\ln\left({\bar p_n-\bar p_j\over \bar p_n+\bar p_j}\right)^2\right\}$$
$$+\sum_{j=1}^{n-1}{4\over \bar p_j}-\sum_{j=n+1}^{N}{4\over \bar p_j},
\quad (n = 1, 2, ..., N). \eqno(4.12)$$
This expression is just the corresponding formula for the short pulse equation [5]. 
Note that if all $\bar p_j$ are real parameters such that $\bar p_j\not=\bar p_k$ for $j\not=k$, then (4.12) gives the
formula for the phase shift resulting from the overtaking collisions of $N$ loop solitons.
\par
\bigskip
\leftline{\it 4.2. Reduction to the sG equation}\par
\medskip
\noindent The reduction to the sG equation is rather straightforward compared to the previous one for the
short pulse equation.
It turns out that the appropriate scaled variables are given by
$$\bar u=u,\qquad \bar x=\epsilon x,\qquad \bar y=\epsilon y,\qquad \bar t={t\over \epsilon},\qquad \bar \tau={\tau\over \epsilon},$$
$$\bar p_j={p_j\over \epsilon},\qquad \bar \xi_{j0}=\xi_{j0},\ (j=1, 2, ..., N). \eqno(4.13)$$
In terms of the variables (4.13), we can recast equation (1.1) to the sG equation $\bar u_{\bar t\bar x}=\sin \bar u$ 
in the limit of $\epsilon \rightarrow 0$.
The parametric solution (2.30) reduces to the usual form of the $N$-soliton solution of the sG equation i.e., 
$\bar u(\bar x,\bar t)=2{\rm i}\, \ln(\bar f^*/\bar f)$ where $\bar f$ is given by (4.5) with the identification 
$\bar y=\bar x, \bar \tau=\bar t$. 
The phase shift is scaled by $\bar \Delta_n=\epsilon\Delta_n$. It is given by the first term on the right-hand side of (4.12),
reproducing the well-known formula derived by the asymptotic analysis of the $N$-soliton solution of the sG equation [7, 8, 10]. \par
\bigskip
\leftline{\bf 5. Conclusion}\par
\bigskip
\noindent  A direct approach employed in this paper constructs various types of soliton solutions such as  single- and
multi-valued kinks, loop solitons and breathers. These elementary solutions are combined to produce a variety of multisoliton
solutions. As  examples, we presented a solution describing the interaction between a soliton and a breather as well
as a 2-breather solution.
As far as solutions are concerned, one can observe that the generalized sG equation has a rich structure compared with
that of the sG equation. We also demonstrated that the generalized sG equation is reduced to both the short pulse  
and sG equations in appropriate
scaling limits.  Another interesting issue is the generalized sG equation (1.1) with $\nu=1$
for which the method of solution still remains open.
One may apply a sequence of nonlinear transformation similar to that used here to obtain solutions. 
Another direction to be worth investigating is the periodic problem.
The exact method of solution used here will work well for constructing periodic solutions of equation (1.1). See [13, 14]
for periodic  solutions of the short pulse equatoin.
These problem will be pursued in  future works. \par

\newpage

\leftline{\bf Appendix A. Proof of Theorem 2.2} \par
\bigskip
\noindent In this appendix, we show that the tau-functions $f$ and $g$ given respectively by (2.35a) and (2.35b) satisfy the bilinear 
equations (2.24).
First, we rewite $f$ and $g$  in terms of determinants.
For the purpose, we use the formula [15]
$$\sum_{\mu=0,1}{\rm exp}\left[\sum_{j=1}^N\mu_j\xi_j
+\sum_{1\le j<k\le N}\mu_j\mu_k\gamma_{jk}\right]
=\lambda_N\,\det\left({\rm e}^{\zeta_j}\delta_{jk}+{2p_j\over p_j+p_k}\right)_{1\leq j,k\leq N}, \eqno(A.1)$$
where
$$\zeta_j=\xi_j+\sum_{\substack{k=1\\(k\not=j)}}^N\gamma_{jk},
\qquad \lambda_N=\exp\left(-\sum_{1\leq j<k\leq N}\gamma_{jk}\right), \eqno(A.2)$$
and $\delta_{jk}$ is Kronecker's delta.
Since  numerical factors multiplied by $f$ and $g$  have no effects on the proof of (2.24), we use the
determinantal expression given by the right-hand of (A.1) instead of the finite sum. Furthermore, we shift the phase factor $\xi_{j0}$ by
$-\sum_{\substack{k=1\\(k\not=j)}}^N\gamma_{jk}$ so that $\zeta_j=\xi_j$. Consequently, we can express $f$ and $g$ by
the following determinants:
$$f=\det\,A\equiv |A|,\qquad A=(a_{jk})_{1\leq j,k\leq N},\qquad a_{jk}={\rm i}{\rm e}^{\xi_j}\delta_{jk}+{2p_j\over p_j+p_k}, \eqno(A.3)$$
$$g=\det\,B\equiv |B|,\qquad B=(b_{jk})_{1\leq j,k\leq N}, 
\qquad b_{jk}={\rm i}\,{1-p_j\over 1+p_j}\,{\rm e}^{\xi_j}\delta_{jk}+{2p_j\over p_j+p_k}.\eqno(A.4)$$
\par
 For later convenience, we  introduce some notations  as well as formulas for determinants.
Matrices and cofactors associated with any $N\times N$
matrix $A=(a_{jk})_{1\leq j,k\leq N}$ are defined  as follows :
$$A({\bf a}; {\bf b})= \left(\begin{matrix}a_{11}&\ldots&a_{1N}&b_1 \cr
              \vdots&\ddots&\vdots&\vdots \cr
              a_{N1}&\ldots &a_{NN}&b_N \cr
              a_1&\ldots &a_N&0\end{matrix}\right),     \eqno(A.5)$$
$$A({\bf a}, {\bf b}; {\bf c}, {\bf d})= \left(\begin{matrix}a_{11}&\ldots&a_{1N}&c_1&d_1 \cr
              \vdots&\ddots&\vdots&\vdots&\vdots \cr
              a_{N1}&\ldots&a_{NN}&c_N&d_N \cr
              a_1&\ldots&a_N&0&0\cr
               b_1&\ldots &b_N&0&0\end{matrix}\right),    \eqno(A.6)$$
$$A_{jk}={\partial |A|\over \partial a_{jk}},  \eqno(A.7)$$
Here, $A_{jk}$ the cofactor of $a_{jk}$ and ${\bf a}, {\bf b}, {\bf c}$ and ${\bf d}$ are $N$-dimensional vectors,
 ${\bf a}=(a_1, a_2, ..., a_N)$, for example.
 The following
formulas are used frequently in the present analysis [16]:
 $$\left|\begin{matrix}a_{11}&\ldots&a_{1N}&x_1\cr
               \vdots&\ddots&\vdots&\vdots\cr
               a_{N1}&\ldots&a_{NN}&x_N\cr
               y_1&\ldots&y_N&z\end{matrix}\right|
=|A|z-\sum^N_{j,k=1}A_{jk}x_jy_k,   \eqno(A.8)$$
$$|A|=\left|\begin{matrix}a_{11}-1&\ldots&a_{1N}-1&1\cr
               \vdots&\ddots&\vdots&\vdots\cr
               a_{N1}-1&\ldots&a_{NN}-1&1\cr
               -1&\ldots&-1&1\end{matrix}\right|, \eqno(A.9)$$
               $$|A({\bf a}+{\bf b};{\bf c}+{\bf d})|=|A({\bf a};{\bf c})|+|A({\bf a};{\bf d})|+|A({\bf b};{\bf c})|+|A({\bf b};{\bf d})|,
               \eqno(A.10)$$
$$|A({\bf a}, {\bf b}; {\bf c}, {\bf d})||A|=|A({\bf a}; {\bf c})||A({\bf b}; {\bf d})|-|A({\bf a}; {\bf d})||A({\bf b}; {\bf c})|, \eqno(A.11)$$
$$\sum_{j,k=1}^N(f_j+g_k)a_{jk}A_{jk}=\sum_{j=1}^N(f_j+g_j)|A|.\eqno(A.12)$$
Formula (A.11) is Jacobi's identity and
 formula (A.12)  follows from the expansion formulas for determinants,
$\sum_{k=1}^Na_{ik}A_{jk}=\delta_{ij}|A|, \sum_{k=1}^Na_{ki}A_{kj}=\delta_{ij}|A|$.  
\par
Let us now proceed to the proof. First, we modify the determinant $|B|$. We extract a factor $2p_j$ from the $j$th row of $|B|$ and then extract a
factor $(1+p_j)^{-1}$ from the $j$th column $(j=1, 2, ..., N)$. Subsequently, the determinant is modified by formula (A.9).
We extract a factor $1-p_j$ from the $j$th row of the resultant determinant and then multiply  the $j$th row by a factor $2p_j$ 
$\ (j=1, 2, ..., N)$. We then find 
$$g=\mu(|A|+2|A(-{\bf 1};{\bf q-1})|)=\mu(|A|+2|A({\bf 1};{\bf 1})|-2|A({\bf 1};{\bf q})|), \eqno(A.13)$$
where
$${\bf q}=\left({1\over 1-p_1}, {1\over 1-p_2}, ..., {1\over 1-p_N}\right),\qquad {\bf 1}=(1, 1, ..., 1), 
\qquad \mu={\prod_{j=1}^N(1-p_j)\over \prod_{j=1}^N(1+p_j)}. \eqno(A.14)$$
The last line of (A.13) is a consequence of formula (A.10).  If we use (A.9), we can rewrite (A.13) as
$$g=\mu(|\bar A|+|\bar A({\bf 1};{\bf 1})|-2|\bar A({\bf 1};{\bf q})|), \eqno(A.15)$$
where $\bar A$ is a skew-Hermitian matrix defined by
$$\bar A=(\bar a_{jk})_{1\leq j,k\leq N},\qquad  \bar a_{jk}={\rm i}{\rm e}^{\xi_j}\delta_{jk}+{p_j-p_k \over p_j+p_k}. \eqno(A.16)$$
Since $|\bar A|^*=(-1)^N|\bar A|$, the comlex conjugate  of $g$ becomes
$$g^*=(-1)^N\mu(|\bar A|-|\bar A({\bf 1};{\bf 1})|+2|\bar A({\bf q};{\bf 1})|). \eqno(A.17)$$
In view of the formulas $|A|=|\bar A|-|\bar A({\bf 1};{\bf 1})|$ and $|\bar A({\bf q};{\bf 1})|=| A({\bf q};{\bf 1})|$ which follow from 
(A.9), (A.17) reduces to
$$g^*=(-1)^N\mu(|A|+2|A({\bf q};{\bf 1})|). \eqno(A.18)$$
Similarly, one has
$$f^*=(-1)^N(|A|+2|A({\bf 1};{\bf 1})|). \eqno(A.19)$$
It follows from (A.13), (A.18) and (A.19) that
$${1\over 2}(fg^*-f^*g)=(-1)^N\mu[(| A({\bf 1};{\bf q})|+| A({\bf q};{\bf 1})|)|A|$$
$$+2(| A({\bf 1};{\bf q})|-|A({\bf 1};{\bf 1})|)|A({\bf 1};{\bf 1})|-2|A||A({\bf 1};{\bf 1})|]. \eqno(A.20)$$
\par
The next step is to calculate the right-hand side of (2.24a). First, applying the differential rule of determinant to $f$, one has
$$f_y=\sum_{j,k=1}^N{\partial a_{jk}\over\partial y}A_{jk}=\sum_{j,k=1}^N(p_j{\rm i}{\rm e}^{\xi_j}\delta_{jk})A_{jk}
={1\over 2}\sum_{j,k=1}^N(p_j+p_k)\left(a_{jk}-{2p_j\over p_j+p_k}\right)A_{jk}. \eqno(A.21)$$
By virtue of (A.8) and (A.12), we can recast (A.21) to
$$f_y=\sum_{j=1}^Np_j|A|+|A({\bf 1};{\bf p})|, \eqno(A.22)$$
where ${\bf p}=(p_1, p_2, ..., p_N)$. A similar calculation leads to 
$$g_y^*=\sum_{j=1}^Np_j|B|^*+(-1)^N\mu(|A({\bf 1};{\bf 1})|+|A({\bf 1};{\bf p})|-|A({\bf q};{\bf 1})|-|A({\bf q};{\bf p})|
-2|A({\bf q}, {\bf 1}; {\bf p}, {\bf 1})|). \eqno(A.23)$$
It follows from (A.18), (A.22) and (A. 23) that
$$D_yg^*\cdot f=(-1)^N\mu[(|A({\bf 1};{\bf 1})|-|A({\bf q};{\bf 1})|-|A({\bf q};{\bf p})|
-2|A({\bf q}, {\bf 1}; {\bf p}, {\bf 1})|)|A|-2|A({\bf 1};{\bf p})||A({\bf q};{\bf 1})|]. \eqno(A.24)$$
Using Jacobi's idenity (A.11) with ${\bf a}={\bf q},\ {\bf b}={\bf 1},\ {\bf c}={\bf p},\ {\bf d}={\bf 1}$, (A.24) simplifies to
$$D_yg^*\cdot f=(-1)^N\mu[(|A({\bf 1};{\bf 1})|-|A({\bf q};{\bf 1})|-|A({\bf q};{\bf p})|)|A|
-2|A({\bf q};{\bf p})||A({\bf 1};{\bf 1})|]. \eqno(A.25)$$
Referring to (A.20) and (A.25), we obtain
$$D_yg^*\cdot f-{1\over 2}(fg^*-f^*g)$$
$$=(-1)^N\mu(-|A({\bf 1};{\bf 1})|+|A({\bf 1};{\bf q)})|-|A({\bf q};{\bf p})|)(|A|+2|A({\bf 1};{\bf 1})|). \eqno(A.26)$$
Let $P=-|A({\bf 1};{\bf 1})|+|A({\bf 1};{\bf q)})|-|A({\bf q};{\bf p})|$. Applying (A.8) with $q_j=1/(1-p_j)$, $P$ becomes
$$P=\sum_{j,k=1}^N(1-q_j+p_jq_k)A_{jk}=\sum_{j,k=1}^N{p_j(p_k-p_j)\over (1-p_j)(1-p_k)}A_{jk}. \eqno(A.27)$$
If we extract a factor $p_l$ from the $l$th row of $A_{jk}\ (l=1, 2, .., N; l\not=j)$, $P$ is modified as
$$P=\left(\prod_{l=1}^Np_l\right)\sum_{j,k=1}^N{p_k-p_j\over (1-p_j)(1-p_k)}\hat A_{jk}, \eqno(A.28)$$
where $\hat A_{jk}$ is the cofactor of the $(j,k)$ element of the matrix $\hat A$ defined by
$${\hat A}=(\hat a_{jk})_{1\leq j,k\leq N}, 
\qquad \hat a_{jk}={{\rm i\rm e}^{\xi_j}\over p_j}\delta_{jk}+{2\over p_j+p_k}. \eqno(A.29)$$
Since $\hat A$ is a symmetric matrix, $\hat A_{jk}=\hat A_{kj}$. Taking this relation into (A.28), we conclude  that $P=0$. This completes the
proof of (2.24a). \par
The proof of (2.24b) follows immediately from that of (2.24a) by a symmetry of the tau-functions.
Indeed, if we exchange the variables $y$ and $\tau$ and subsequently replace the parameter $p_j$ by
$p_j^{-1}\ (j=1, 2, ..., N)$, we then see that $f$ from (A.3) is unchanged whereas $g$ from (A.4) is transformed to $g^*$.
Thus, under this manipulation, the bilinear equation (2.24b) turns out to the bilinear equation (2.24a),
which completes the proof of (2.24b). \par
\newpage
\leftline{\bf Appendix B. An alternative derivation of the 1-soliton solutions} \par
\bigskip
\noindent The 1-soliton solutions take the form of traveling wave
$$u=u(X), \qquad X=x+c_1t+x_0. \eqno(B.1)$$
Substituting this expression into equation (1.1) with $\nu=-1$ and integrating the resultant ordinary
differential equation  once with
respect to $X$  under the boundary condition $u(-\infty)=0\ ({\rm mod}\, 2\pi)$, we obtain
$$u_X^2={(c_1+1)^2-(\cos u+c_1)^2\over (\cos u+c_1)^2}. \eqno(B.2)$$
Since $u_X^2\geq 0$, we must require that the right-hand side of (B.2) is nonnegative which imposes the condition on
possible values of $c_1$. One can see that this condition becomes $c_1\geq-\cos^2(u/2)$.
To proceed, we define a new variable $\xi$ by
$$X=\int(\cos u +c_1)d\xi. \eqno(B.3)$$
Then,  equation (B.2) reduces to
$$u_\xi=\pm\sqrt{(c_1+1)^2-(\cos u+c_1)^2}. \eqno(B.4)$$
In accordance with values of $c_1$, the solutions are classified to several types, which we shall now detail. \par
\noindent{$\it 1.\ c_1>0$}\par
\noindent In the case of $c_1>0$, (B.4) is integrated through the change of the variable  by $s=\tan(u/2)$.
After an elementary calculation, we obtain
$$s=\pm\sqrt{c_1+1\over c_1}{1\over \sinh\sqrt{c_1+1}\xi} \eqno(B.5)$$
and
$$\cos u={1-s^2\over 1+s^2}=1-{2(c_1+1)\over c_1\, \sinh^2\sqrt{c_1+1}\,\xi+c_1+1}. \eqno(B.6)$$
Substituting (B.6) into (B.3) and performing the integration with respect to $\xi$, we find
$$X=(c_1+1)\xi-{\rm ln}\,{\sqrt{c_1+1}+\tanh\sqrt{c_1+1}\,\xi \over \sqrt{c_1+1}-\tanh\sqrt{c_1+1}\,\xi}+\xi_0, \eqno(B.7)$$
where $\xi_0$ is an integration constant. It follows from (B.5) and the boundary condition for $u$ that
$$u=2\,\tan^{-1}\left(\sqrt{c_1\over c_1+1}\sinh\sqrt{c_1+1}\,\xi\right) +\pi. \eqno(B.8)$$
If we put
$$c_1={1\over p_1^2}-1\ (0<p_1<1), \qquad \xi=p_1(\xi_1-d_1),\qquad d_1=\tanh^{-1}p_1, $$
$$\xi_0={\rm ln}\left({1-p_1\over 1+p_1}\right)+{d_1\over p_1}+y_0, \eqno(B.9)$$
we can see that (B.7) and (B.8) coincide with (3.4) and (3.3a), respectively. \par
\medskip
\noindent{$\it 2.\ -1< c_1<0$}\par
\noindent In this case, a calculation similar to case 1 gives
$$s=\pm\sqrt{c_1+1\over -c_1}{1\over \cosh\sqrt{c_1+1}\,\xi} \eqno(B.10)$$
and
$$u=-2\,\tan^{-1}\left(\sqrt{-c_1\over c_1+1}\cosh\sqrt{c_1+1}\,\xi\right) +\pi, \eqno(B.11)$$
$$X=(c_1+1)\xi-{\rm ln}\,{1+\sqrt{c_1+1}\,\tanh\sqrt{c_1+1}\,\xi \over 1-\sqrt{c_1+1}\,\tanh\sqrt{c_1+1}\,\xi}+\xi_0. \eqno(B.12)$$
If we put
$$c_1={1\over p_1^2}-1\ (p_1>1), \qquad \xi=p_1(\xi_1-d_1),\qquad d_1=\tanh^{-1}{1\over p_1}, $$
$$\xi_0={\rm ln}\left({p_1-1\over p_1+1}\right)+{d_1\over p_1}+y_0, \eqno(B.13)$$
we can reproduce the parametric solution (3.3a) and (3.4).\par
\medskip
\noindent {$\it 3.\ c_1=0$}\par
\noindent For the special value $c_1=0$, integration of (B.3) and (B.4) can be
performed readily, giving rise to the solution
$$u={\pi\over 2}+\tan^{-1}(\sinh\,\xi), \eqno(B.14)$$
$$X=-{\rm ln}(\cosh\,\xi) +\xi_0. \eqno(B.15)$$
It is easy to confirm that (B.14) and (B.15) coinside with (3.12a) and (3.12b), respectively. \par

\newpage
\leftline{\bf References}\par
\begin{enumerate}[{[1]}]
\item Fokas AS 1995 {\it On a class of physically important integrable equations} {\it Phys. D} {\bf 87} 145
\item Lenells L and Fokas AS 2009 {\it On a novel integrable generalization of the sine-Gordon equation} arXiv: 0909.2590v1[nlin. SI]
\item Hirota R 1980 {\it Direct Methods in Soliton Theory}
in {\it Solitons} ed RK Bullough and DJ Caudrey 
Topics in Current Physics Vol. 17 (New York: Springer) p 157
\item Matsuno Y 1984 {\it Bilinear Transformation Method} (New York: Academic Press)
\item Matsuno Y 2007 {\it Multiloop soliton and multibreather solutions of the short pulse model equation}
  {\it J. Phys. Soc. Japan} {\bf 76} 084003
\item Lamb, Jr GL 1980 {\it Elements of Soliton Theory} (New York: John Wiley \& Sons)
\item Hirota R 1972 {\it Exact solution of the sine-Gordon equation for multiple collisions of solitons}
    {\it J. Phys. Soc. Japan} {\bf 33} 1459
\item Caudrey RJ, Gibbon JD, Eilbeck JC and Bullough RK 1973 {\it Exact multisoliton solutions of the self-induced
transparency and sine-Gordon equation} {\it Phys. Rev. Lett.}  {\bf 30} 237
\item Ablowitz MJ, Kaup DJ, Newell AC and Segur H 1973 {\it Method for solving the sine-Gordon equation} 
{\it Phys. Rev. Lett.}  {\bf 30} 1262
\item Takhtadzhyan LA 1974 {\it Exact theory of propagation of ultrashort optical pulses in two-level media}
{\it Soviet Phys. JETP} {\bf 39} 228
\item Sh\"affer T and Wayne CE 2004 {\it Propagation of ultra-short optical pulses in cubic nonlinear media}
{\it Phys. D} {\bf 196} 90
\item Matsuno Y 2006 {\it Cusp and loop soliton solutions of short-wave models for the Camassa-Holm and Degasperis-Procesi equations}
{\it Phys. Lett. A} {\bf 359} 451
\item Matsuno Y 2008 {\it Periodic solutions of the short pulse model equation} {\it J. Math. Phys.} {\bf 49} 073508
\item Matsuno Y 2009 {\it Soliton and periodic solutions of the short pulse model equation} 
in {\it Handbook of Solitons: Research, Technology and Applications} ed SP Lang and SH Bedore (New York: Nova Publishers) to appear
\item Matsuno Y 2000 {\it Multiperiodic and multisoliton solutions of a nonlocal nonlinear Schr\"odinger equation
for envelope waves} {\it Phys. Lett. A} {\bf 278} 53
\item Vein R and Dale P 1999 {\it Determinants and Their Applications in Mathematical Physics} (New York: Springer) 
\end{enumerate}

\newpage
\leftline{\bf Figure captions} \par
\begin{itemize}
\item[{\bf Figure 1.}] The profile of a regular kink $u$ (solid line) and corresponding profile of $v\equiv u_X$ (broken line). 
The parameter $p_1$ is set to $0.4$ and the parameter $y_0$ is chosen such that the center position of $u_X$ is at $X=0$. 
 
\item[{\bf Figure 2.}] The profile of a singular kink with the parameter $p_1=0.9$. 

\item[{\bf Figure 3.}] The profile of a loop soliton with the parameter $p_1=2.0$.

\item[{\bf Figure 4.}] The profile of a stationary solution with the parameter $p_1=1.0$.

\item[{\bf Figure 5.}]  The profile of a kink-kink solution $u$ (solid line) 
and corresponding profile of $v\equiv u_x$ (broken line) for three different times, a: $t=0$,\ b: $t=2$,\ c: $t=3$.
The parameters are chosen as $p_1=0.3,\ p_2=0.6,\ \xi_{10}=-5,\ \xi_{20}=0$.

\item[{\bf Figure 6.}]  The profile of a kink-loop soliton solution $u$  
 for three different times, a: $t=0$,\ b: $t=4$,\ c: $t=8$.
The parameters are chosen as $p_1=0.5,\ p_2=2.0,\ \xi_{10}=0, \xi_{20}=25$.

\item[{\bf Figure 7.}]  The profile of a breather solution $v\equiv u_x$  
 for three different times, a: $t=0$, \ b: $t=5$,\ c: $t=10$.
The parameters are chosen as $p_1=0.2+0.5\,{\rm i},\ p_2=p_1^*=0.2-0.5\,{\rm i},\ \xi_{10}=\xi_{20}^*=0$.

\item[{\bf Figure 8.}]  The profile of a soliton-breather solution $v\equiv u_x$  
 for three different times, a: $t=0$, \ b: $t=15$, c: $t=25$.
The parameters are chosen as $p_1=0.2+0.5\,{\rm i},\ p_2=p_1^*=0.2-0.5\,{\rm i}, \ p_3=0.3,\ \xi_{10}=\xi_{20}=0,\ \xi_{30}=-30$.

\item[{\bf Figure 9.}]  The profile of a breather-breather solution $v\equiv u_x$  
 for three different times, a: $t=0$, \ b: $t=40$, c: $t=70$.
The parameters are chosen as $p_1=0.1+0.5\,{\rm i}, p_2=0.16+0.8\,{\rm i}, p_3=p_1^*=0.1-0.5\,{\rm i}, 
\ p_4=p_2^*=0.16-0.8\,{\rm i}, \ \xi_{10}=\xi_{30}^*=-10,
\ \xi_{20}=\xi_{40}^*=0$. 
\end{itemize}

\newpage
\begin{center}
\includegraphics[width=6cm]{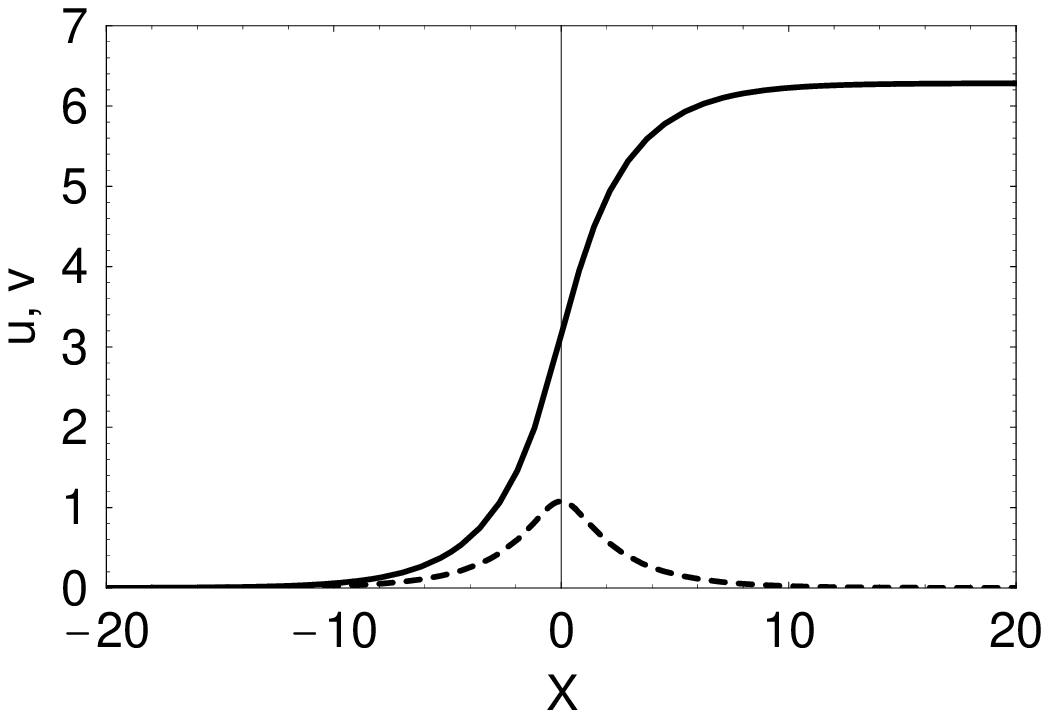}
\end{center}
\centerline{\bf Figure 1}

\bigskip
\begin{center}
\includegraphics[width=6cm]{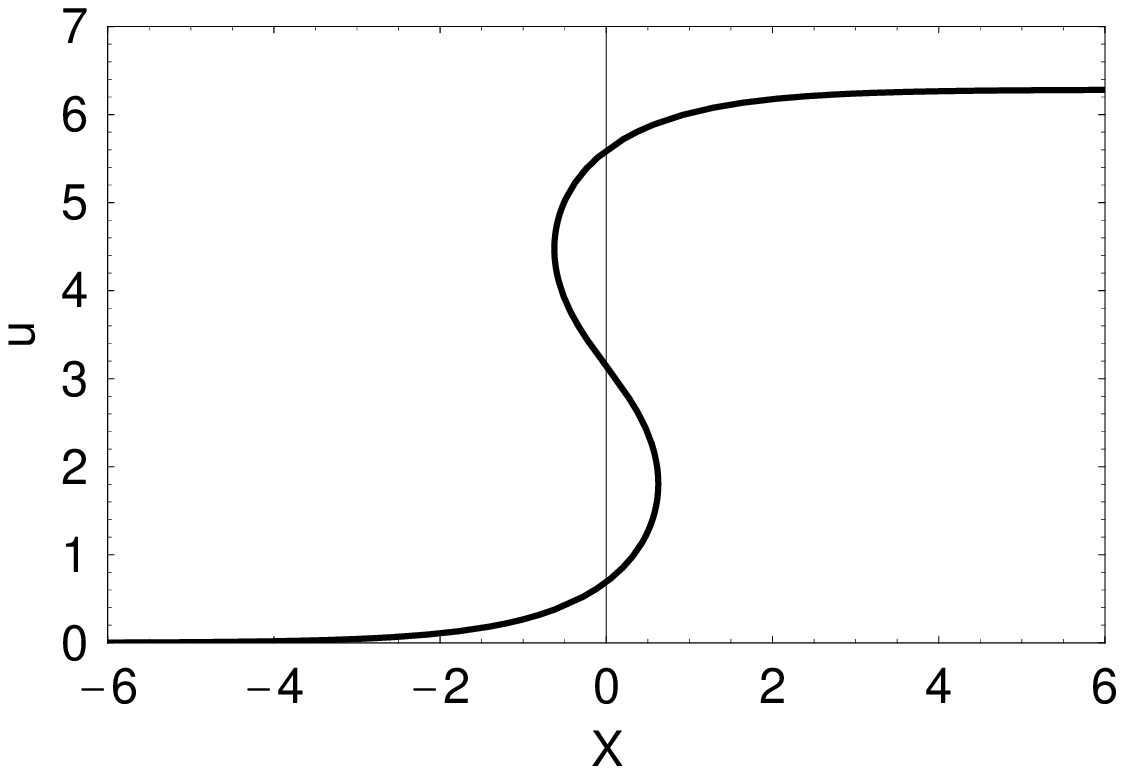}
\end{center}
\centerline{\bf Figure 2}
\bigskip
\begin{center}
\includegraphics[width=6cm]{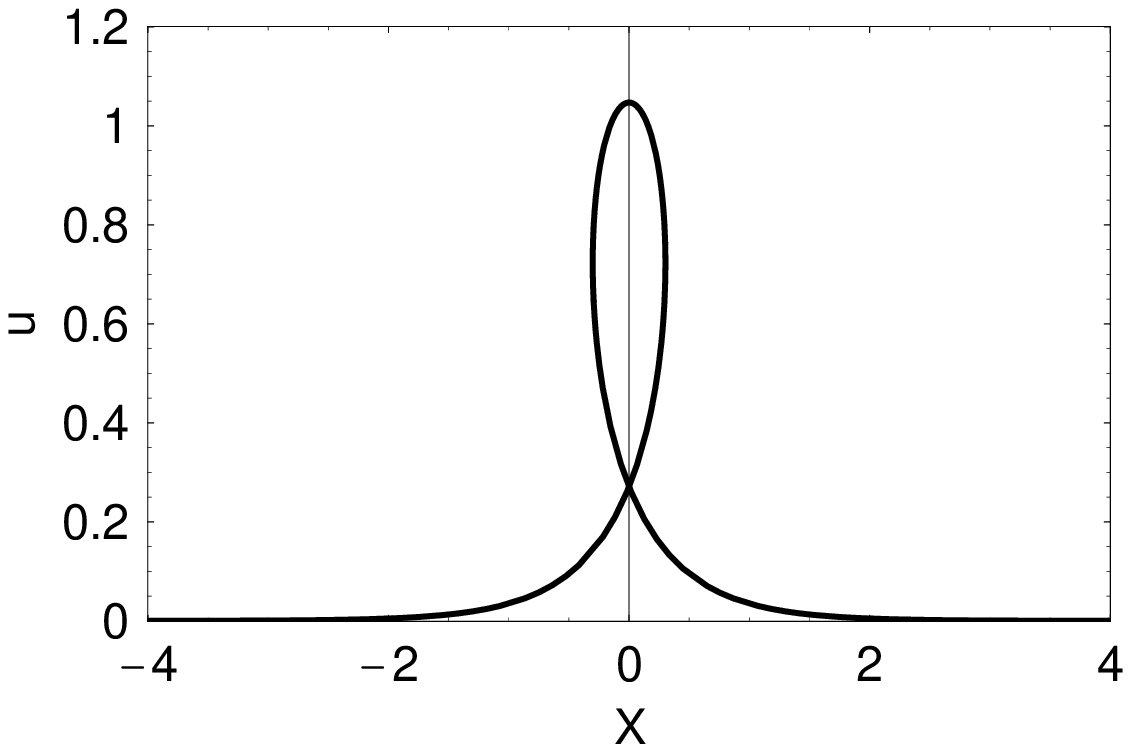}
\end{center}
\centerline{\bf Figure 3}
\bigskip
\begin{center}
\includegraphics[width=6cm]{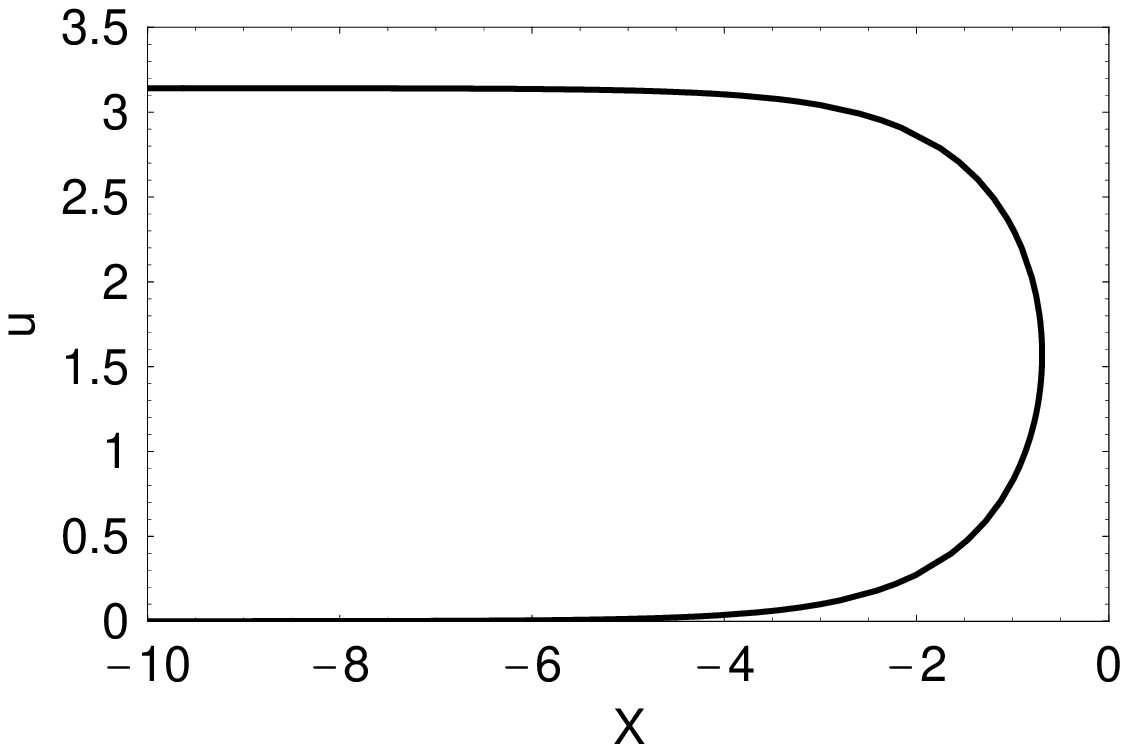}
\end{center}
\centerline{\bf Figure 4}

\newpage
\begin{center}
\includegraphics[width=10cm]{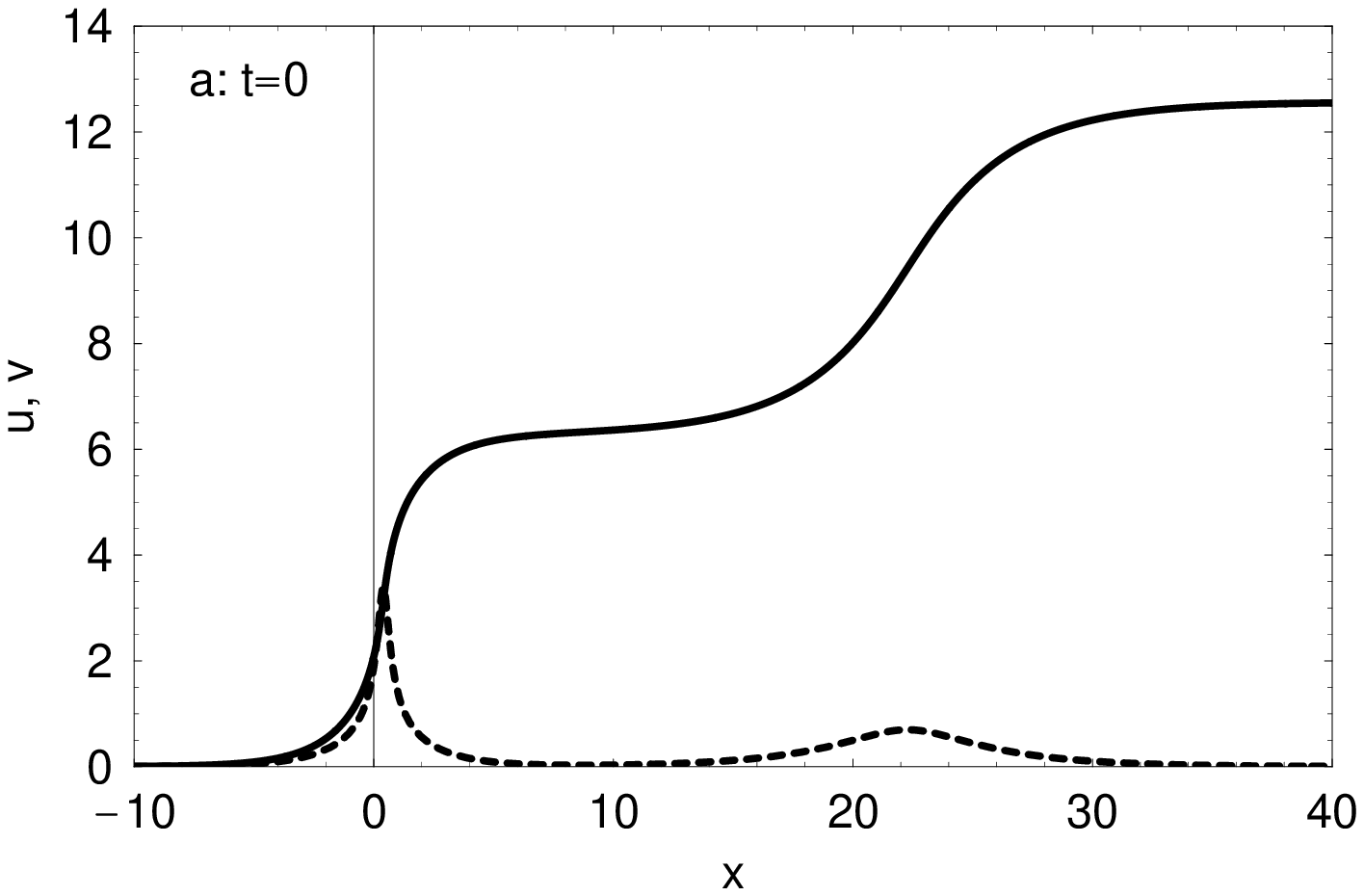}
\end{center}
\bigskip
\begin{center}
\includegraphics[width=10cm]{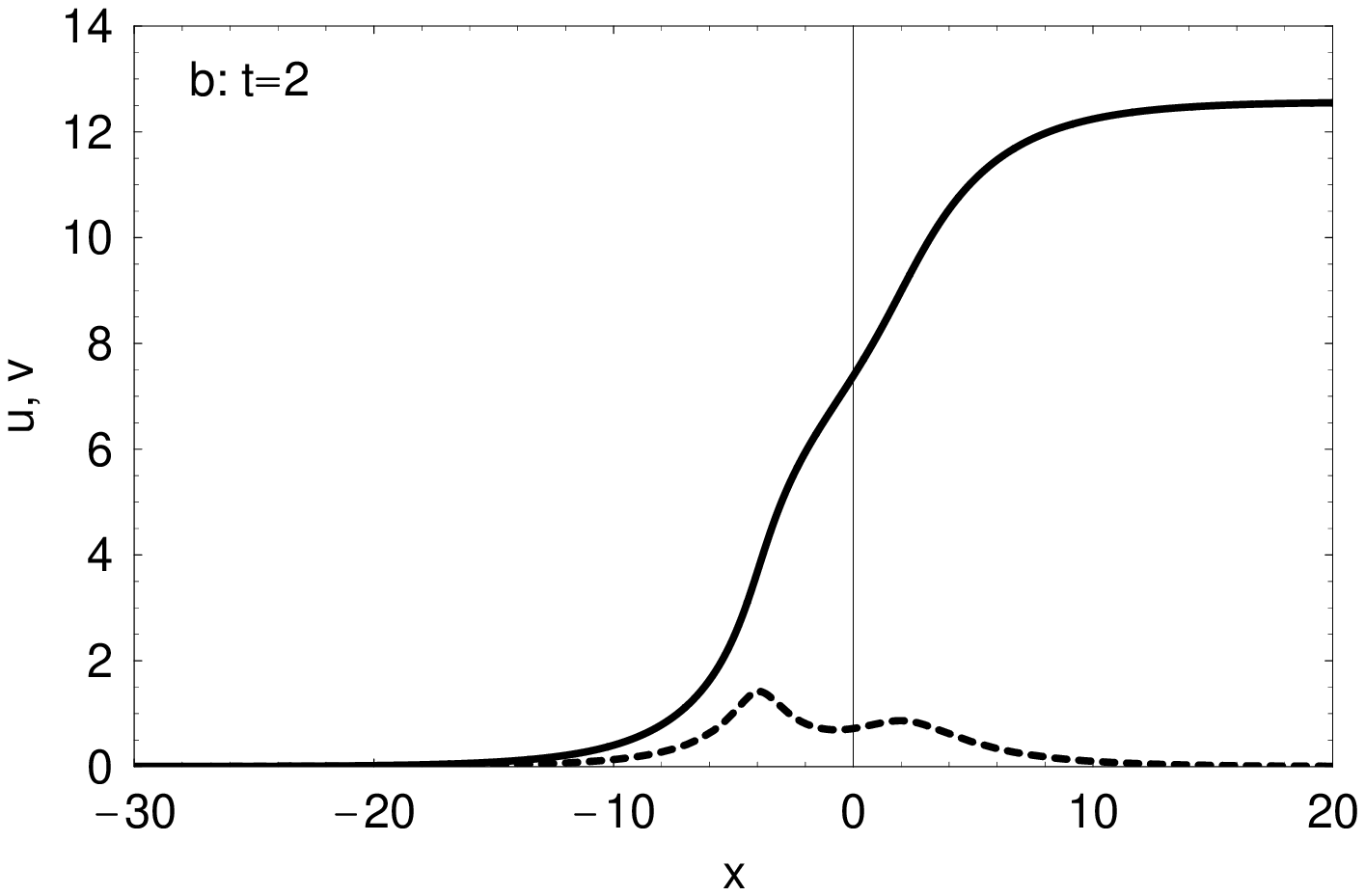}
\end{center}
\bigskip
\begin{center}
\includegraphics[width=10cm]{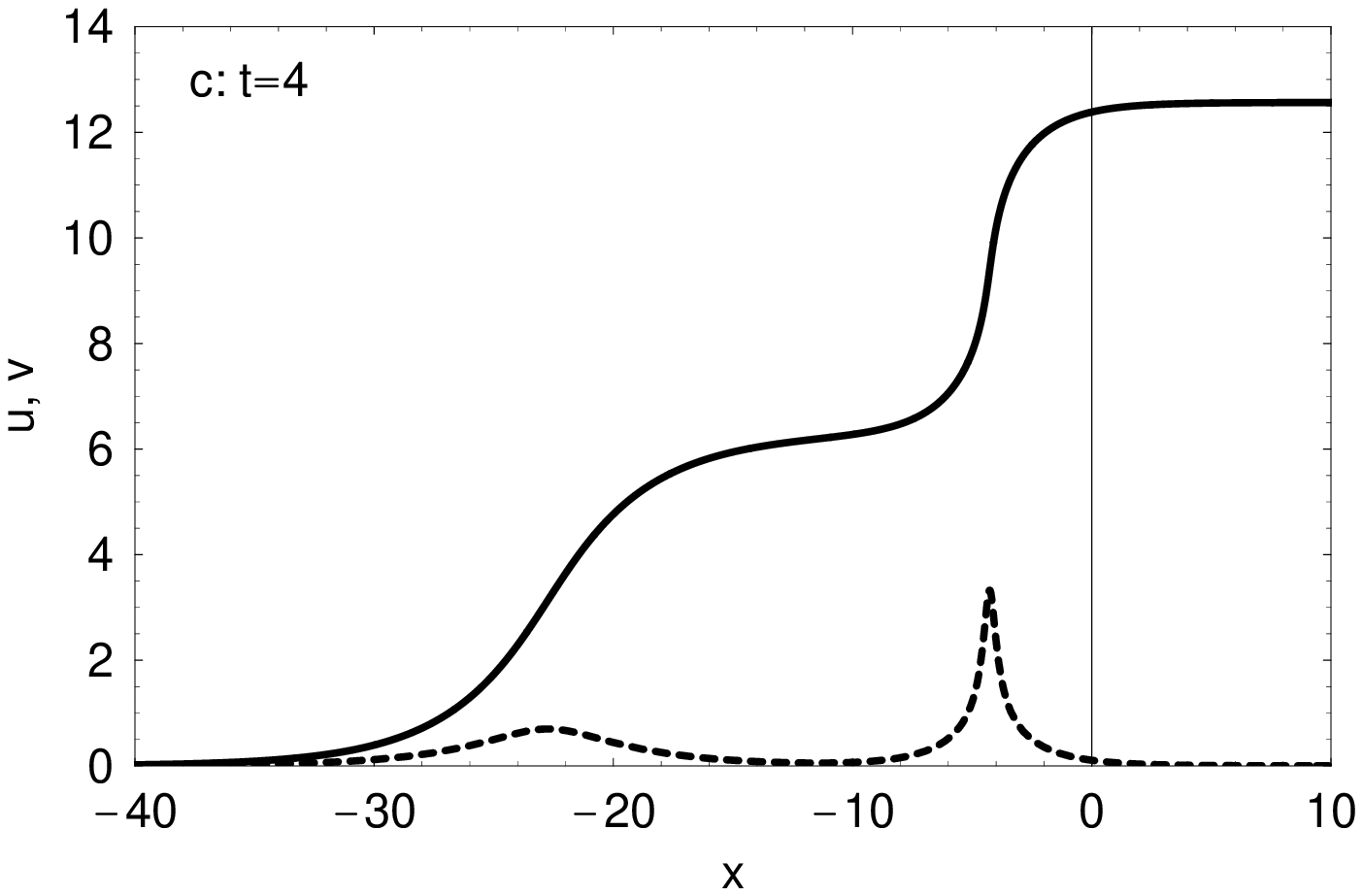}
\end{center}
\centerline{\bf Figure 5 a-c}

\newpage
\begin{center}
\includegraphics[width=10cm]{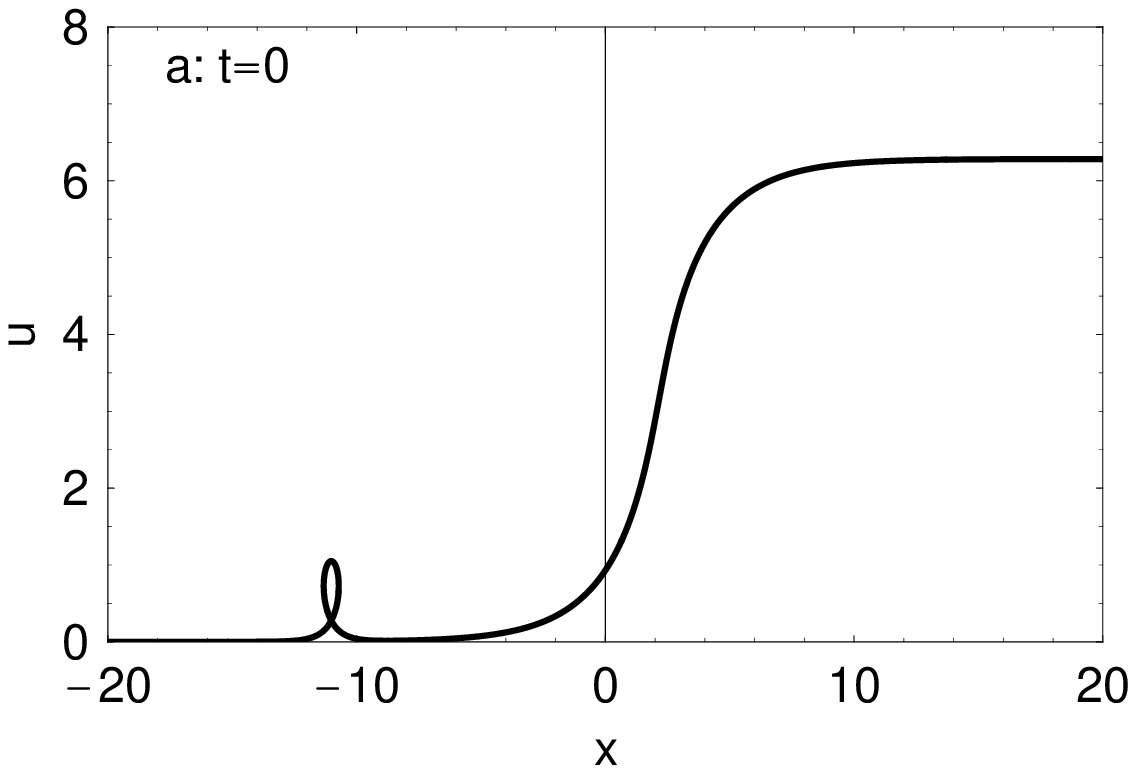}
\end{center}
\bigskip
\begin{center}
\includegraphics[width=10cm]{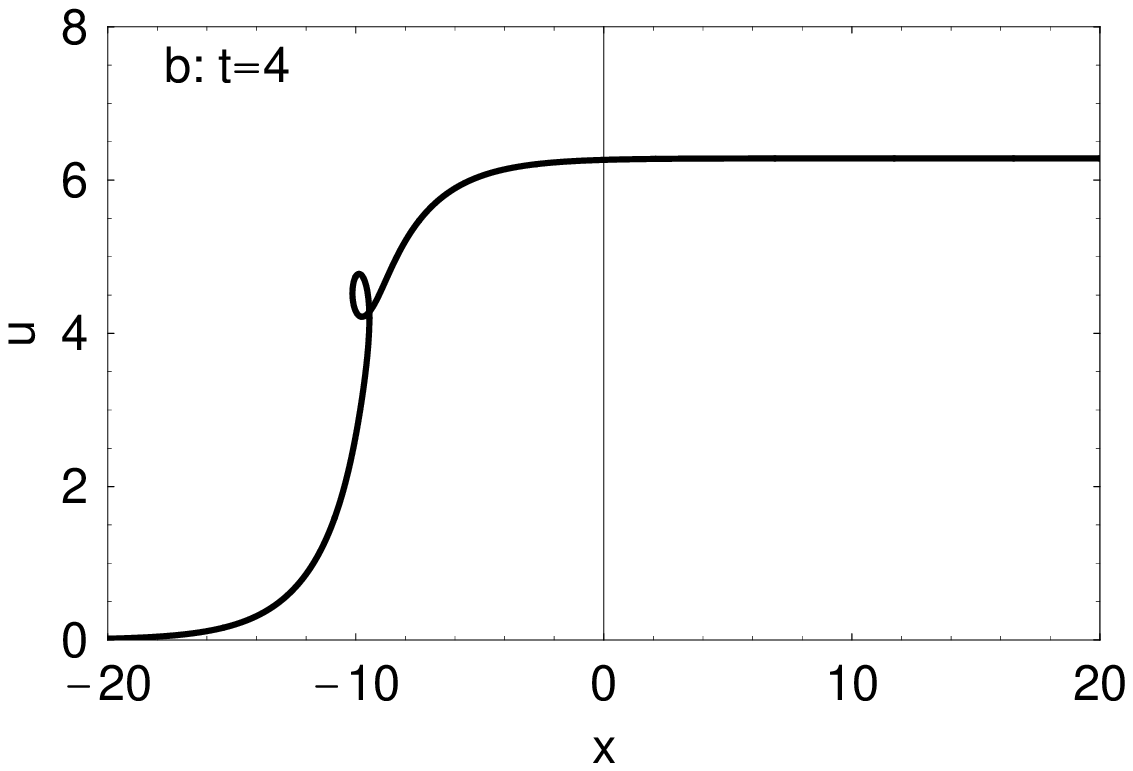}
\end{center}
\bigskip
\begin{center}
\includegraphics[width=10cm]{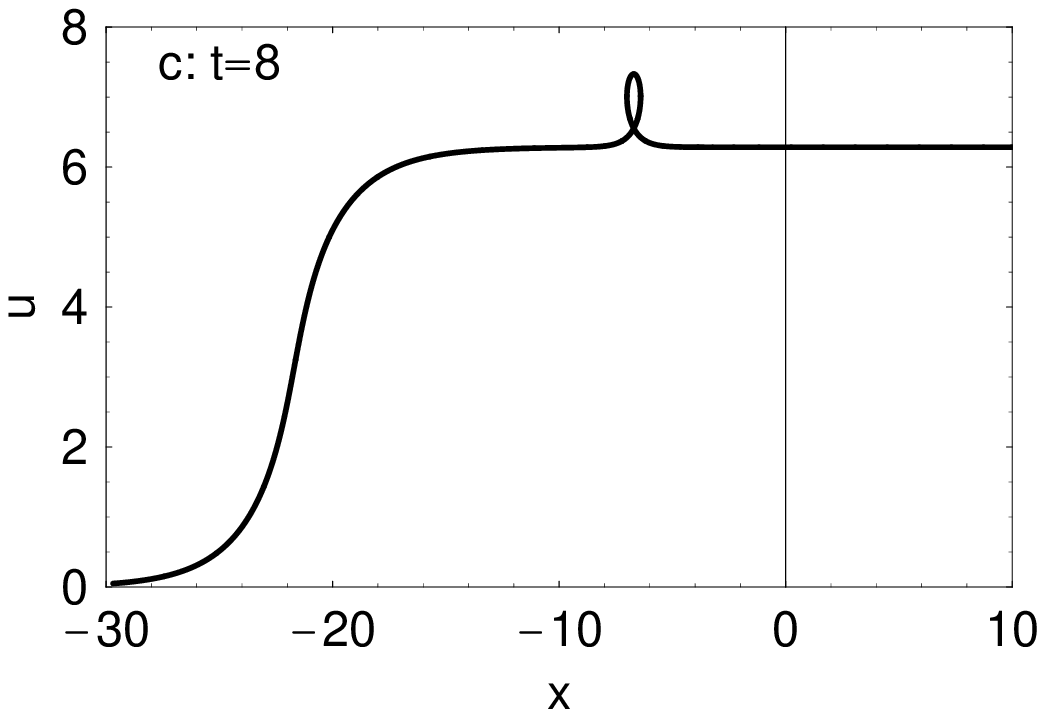}
\end{center}
\centerline{\bf Figure 6 a-c}

\newpage
\begin{center}
\includegraphics[width=10cm]{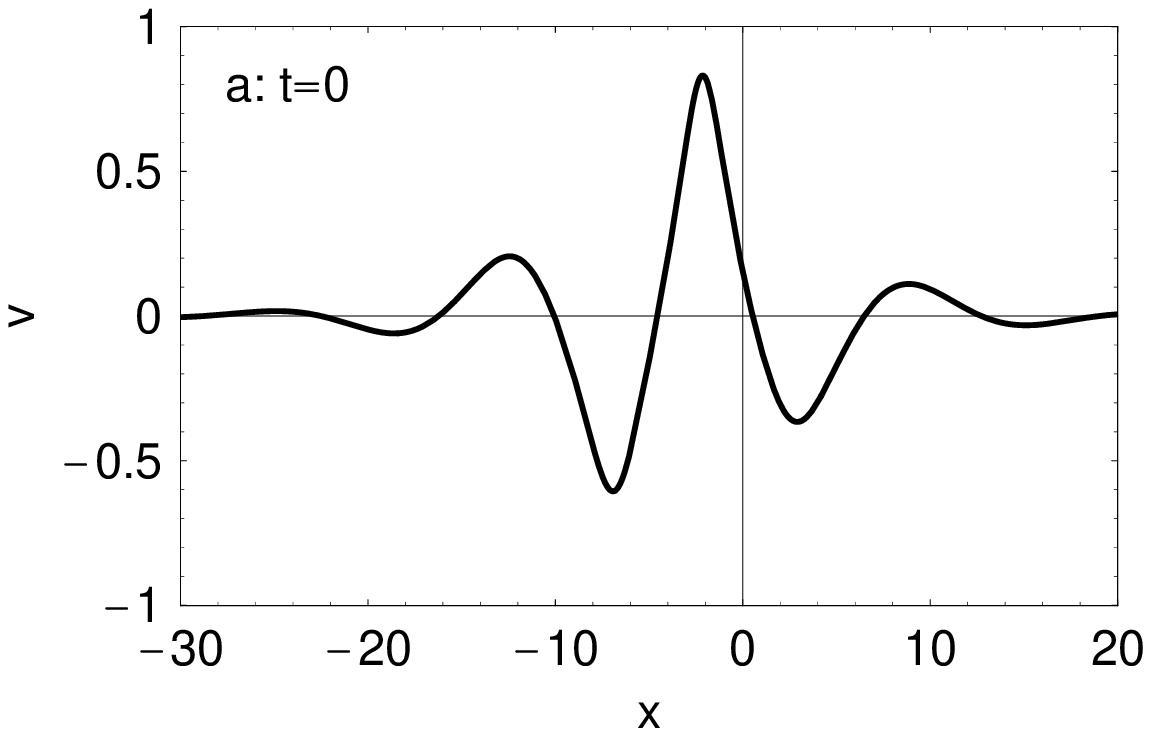}
\end{center}
\bigskip
\begin{center}
\includegraphics[width=10cm]{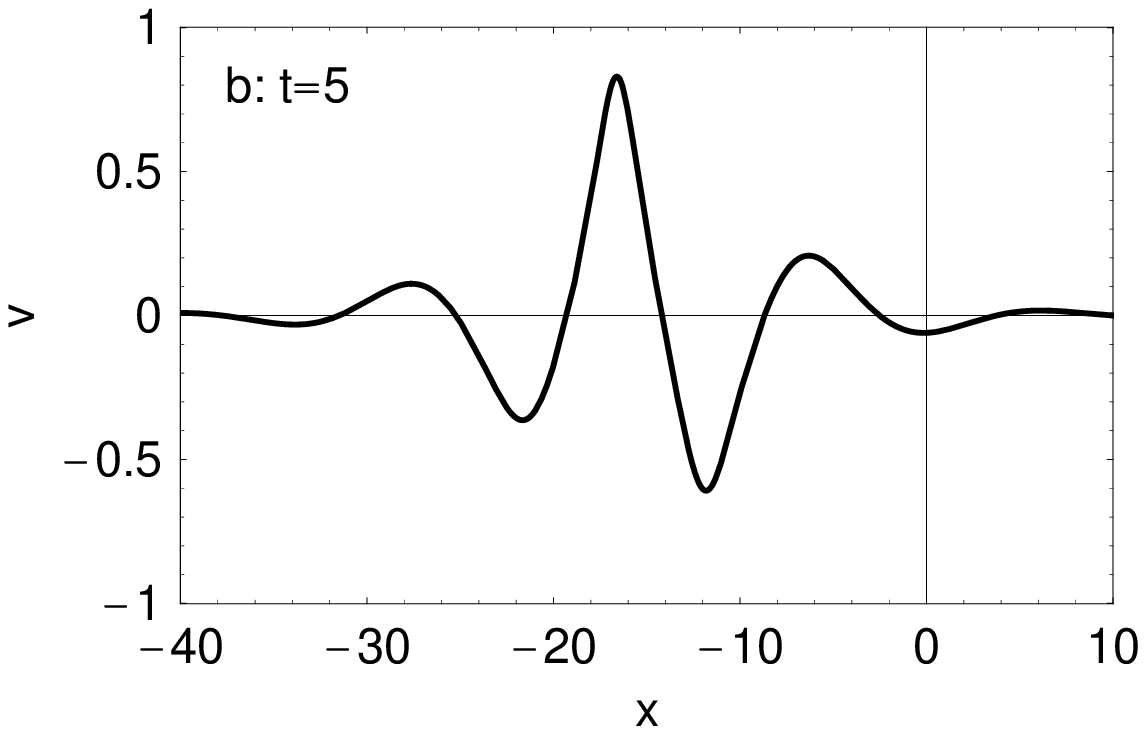}
\end{center}
\bigskip
\begin{center}
\includegraphics[width=10cm]{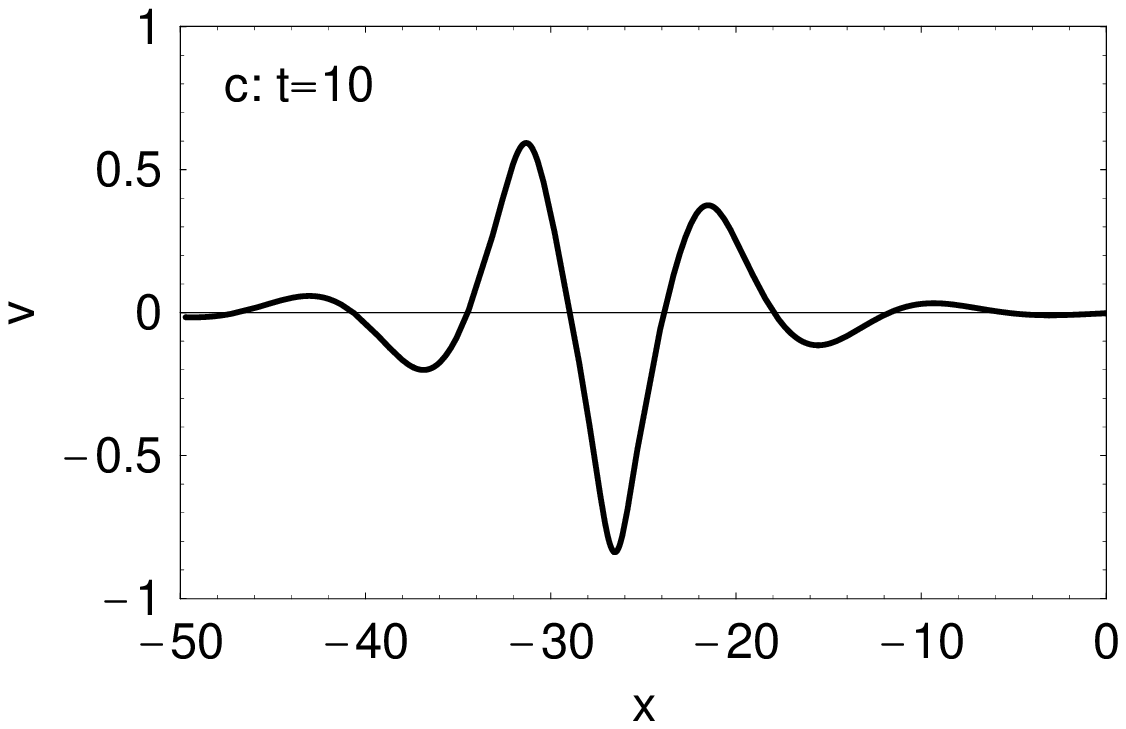}
\end{center}
\centerline{\bf Figure 7 a-c}

\newpage
\begin{center}
\includegraphics[width=10cm]{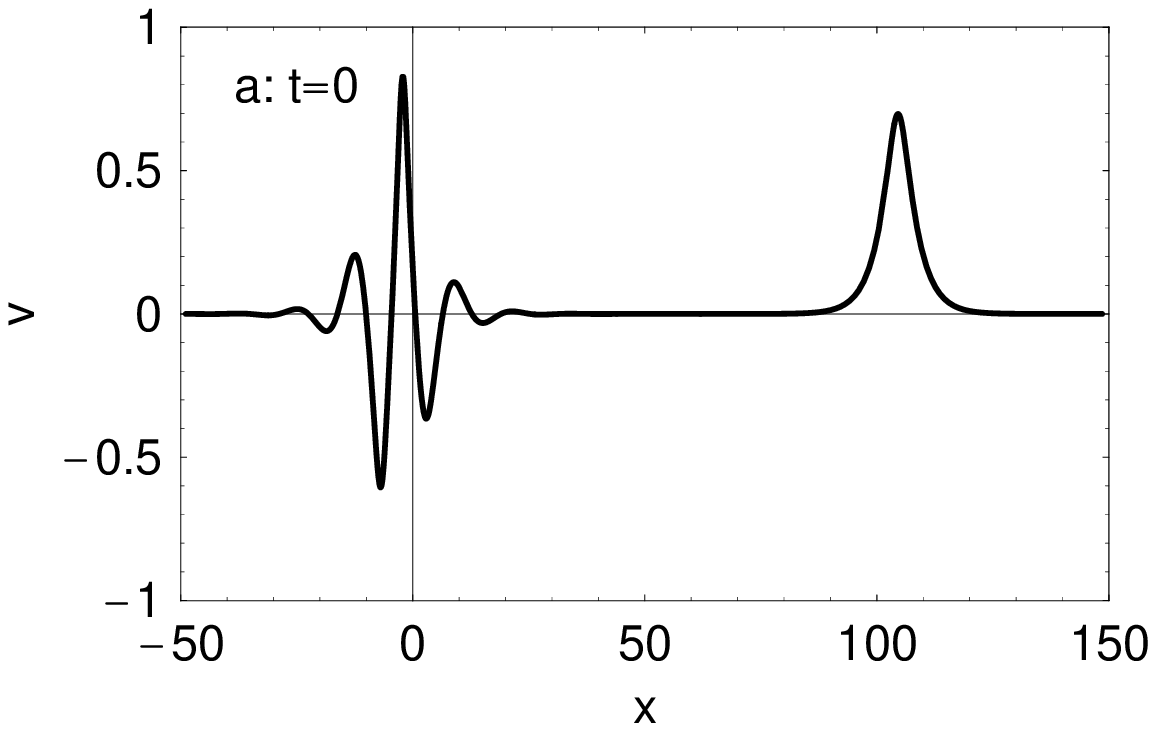}
\end{center}
\bigskip
\begin{center}
\includegraphics[width=10cm]{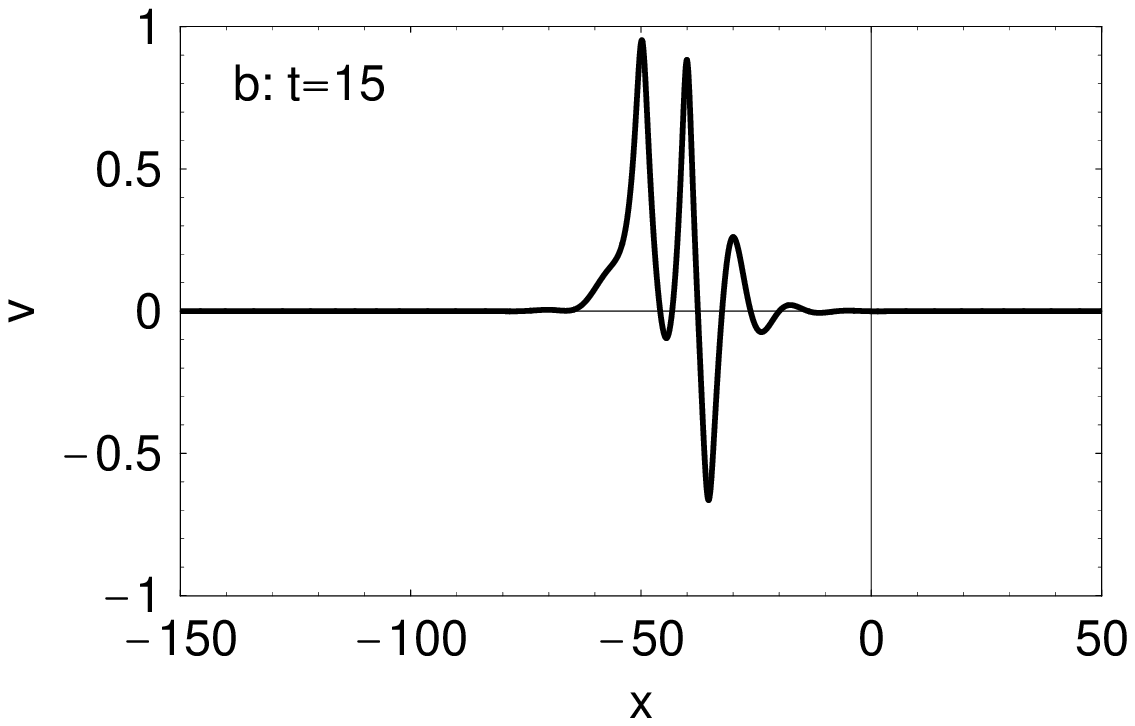}
\end{center}
\bigskip
\begin{center}
\includegraphics[width=10cm]{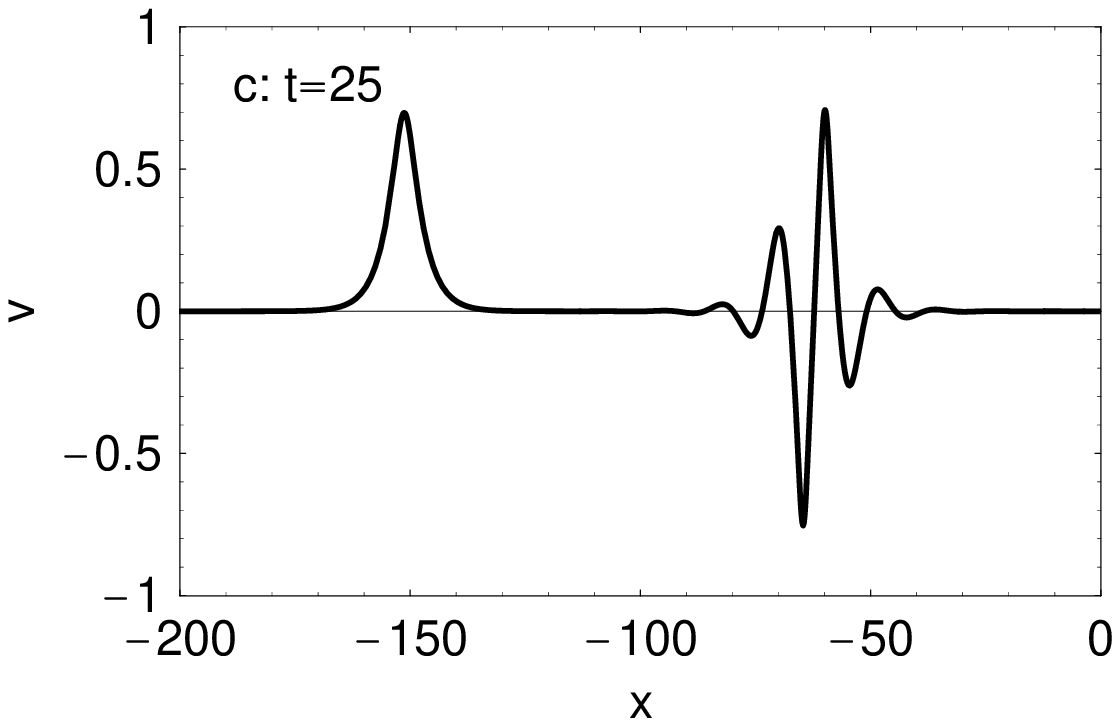}
\end{center}
\centerline{\bf Figure 8 a-c}

\newpage
\begin{center}
\includegraphics[width=10cm]{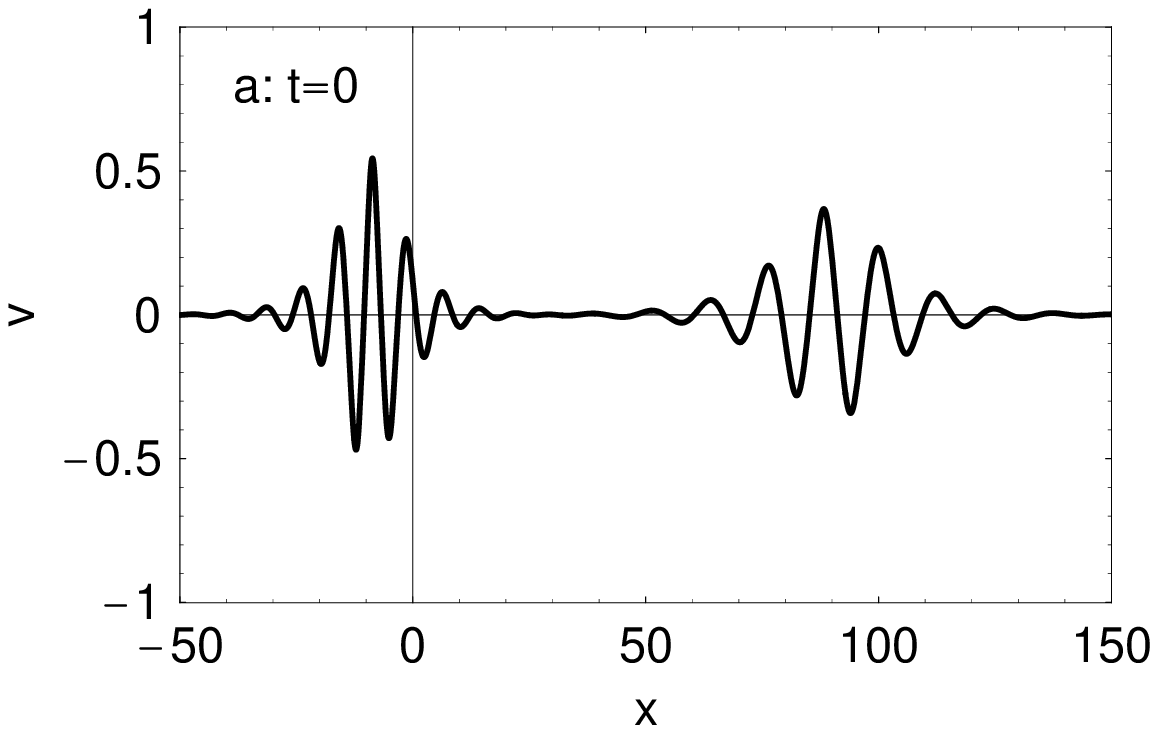}
\end{center}
\bigskip
\begin{center}
\includegraphics[width=10cm]{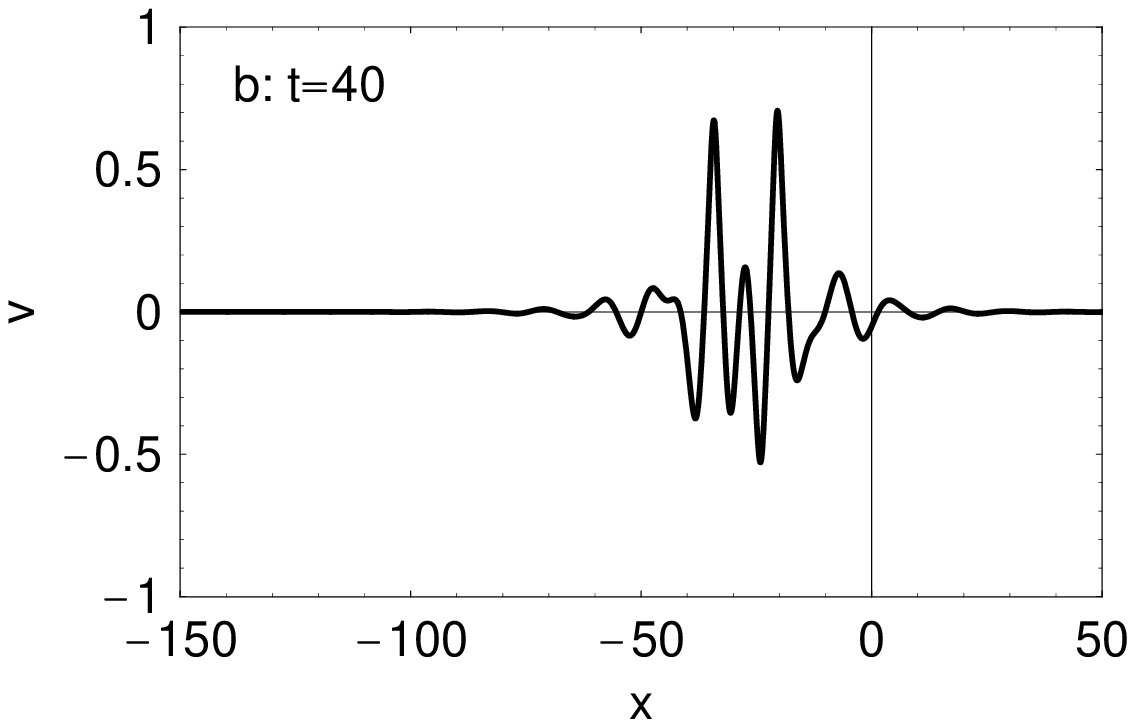}
\end{center}
\bigskip
\begin{center}
\includegraphics[width=10cm]{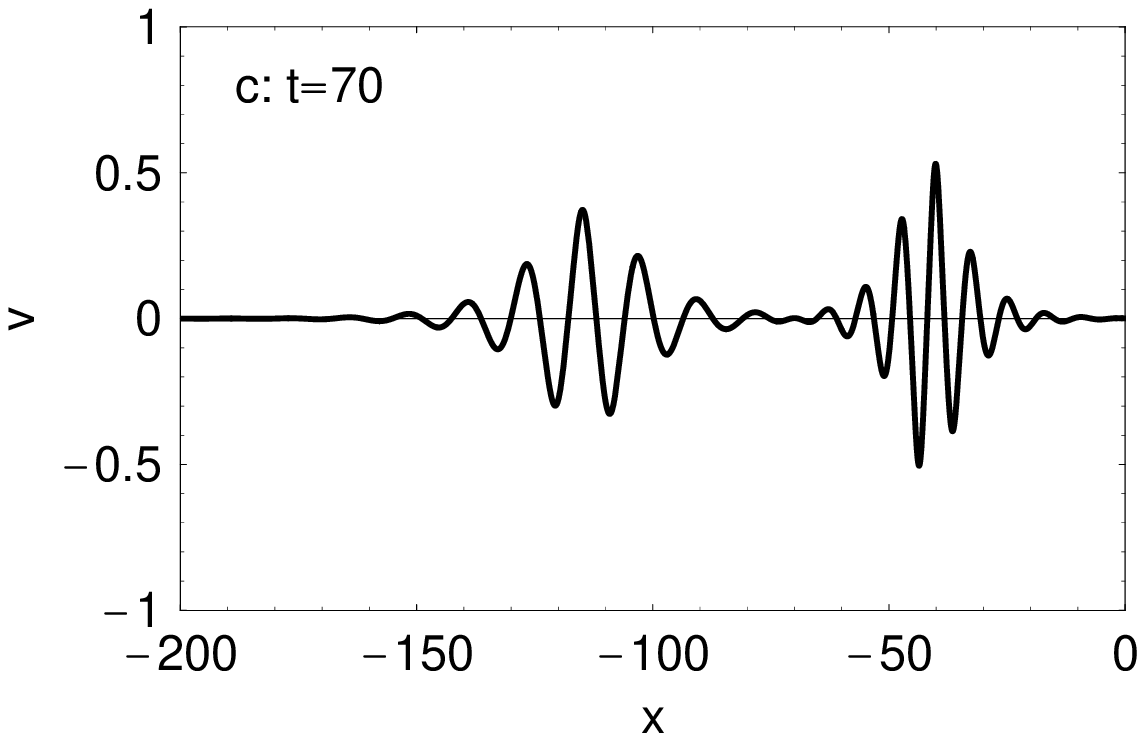}
\end{center}
\centerline{\bf Figure 9 a-c}
\end{document}